# Frequently Used References For Atomic Data In X-ray Spectroscopy

N. Hell[1], G.V. Brown[1], M.E. Eckart[1], A. Fairchild[1], C.A. Kilbourne[2], M.A. Leutenegger[2], F.S. Porter[2], M.C. Witthoeft[2]

[1]Lawrence Livermore National Laboratory, 7000 East Ave., Livermore, CA 94550, USA
[2] NASA Goddard Space Flight Center, 8800 Greenbelt Rd., Greenbelt, MD 20771, USA
contact: hell1@llnl.gov

## Abstract

Accurate atomic physics reference data are a crucial requirement for analysis and interpretation of observed spectra, even more so for observations with high spectral resolution. This document provides a curated list of atomic physics references frequently used for plasma diagnostics in (astrophysical) X-ray spectroscopy, outside of comprehensive plasma models that typically come with their own underlying atomic databases. The list includes references to physical constants, laboratory benchmarks, transition energies, position and line shapes of neutral fluorescence lines, radiative branching ratios, and commonly used notation for prominent transitions. This document also provides quick-look tables for transition energies in H-, He-, and Li-like ions as well as line positions and shapes for fluorescence lines in neutral material. The main focus is on K-shell transitions. For the H- and He-like tables, we cite state-of-the art calculations that we consider currently the best available reference energies. Those energies are considered high accuracy and thus typically used for energy scale calibration in laboratory measurements. Omissions of energy values for the listed transitions in these tables are due to the lack of availability in the chosen references, and are not a statement about the relevance of these lines. Due to their complex and highly source-dependent line shape, the atomic data for fluorescence in neutrals is necessarily of lower accuracy than that for the highly charged ions, and the best reference data for these line shapes typically consist of empirical models derived from very high-resolution laboratory measurements. The table for neutrals provided here is consistent with the reference used for the energy gain scale calibration of XRISM/Resolve.

This document is meant to serve as a resource to help find relevant references and conveniently formatted overview tables. When making use of the information found in these papers, credit should be given to their original authors by citing the appropriate references.

Work by LLNL was performed under the auspices of the US DOE under contract DE-AC52-07NA27344 and supported by NASA grants to LLNL and NASA/GSFC.

## Physical constants

All values from CODATA 2022 (CODATA is updated every 4 years):

CODATA Recommended Values of the Fundamental Physical Constants: 2022
Mohr et al. 2024, arXiv:2409.03787 — doi:10.48550/arXiv.2409.03787
https://ui.adsabs.harvard.edu/abs/2024arXiv240903787M
Quicklook table: https://physics.nist.gov/cuu/pdf/wall_2022.pdf

**Wavelengths ↔ energy conversion: $E\lambda = 12398.41984\,\mathrm{eV\,Å}$**

This is a "defined" value as of 2018, since all involved constants ($h$, $c$, $e$) are defined and therefore considered exact. But note that this value has changed historically (see Hell 2017, p.226, for an overview), which may need to be considered when reading older papers. Also note that the X-ray Data Booklet 2009 has a typo in the value for $e$. The conversion function `_A()` in ISIS uses CODATA 1998.

**Energy ↔ energy conversion: 1 Ry = 0.5 a.u. = 13.605693122990 eV**

**Transition rate ↔ natural line width conversion: $\Delta E/A = \hbar$ with $\hbar = 6.582119569 \cdot 10^{-16}\,\mathrm{eV\,s}$**

where the line width $\Delta E$ is the Lorentzian FWHM in eV and $A$ is in $\mathrm{s}^{-1}$. The Planck constant $h = 2\pi\hbar$ has a defined value as of 2018.

**Doppler broadening: $kT_i = m_i c^2 (\Delta E/E)^2/8/\ln 2$** with $\Delta E$ the Gaussian FWHM

**Electron mass: $m_\mathrm{e}c^2 = 510.9989595069\,\mathrm{keV}$**

**Atomic mass unit: $m_\mathrm{u}c^2 = 931.49410372\,\mathrm{keV} = 1822.888454839255\, m_\mathrm{e}c^2$**

**Speed of light in vacuum: $c = 299\,792\,458\,\mathrm{m\,s}^{-1}$** (exact as of 1973)

## Laboratory benchmarks:

**Fe direct excitation (K$\alpha$)**

EBIT measurement, does not include DR lines (F- to He-like Fe XVIII–XXV):
Decaux et al. 1997, ApJ 482, 1076 — doi:10.1086/304169
https://ui.adsabs.harvard.edu/abs/1997ApJ...482.1076D

**Fe DR (KLL)**

EBIT measurement of KLL for He-like → Li-like Fe DR:
Beiersdorfer et al. 1992, PRA 46, 3812 — doi:10.1103/PhysRevA.46.3812
https://ui.adsabs.harvard.edu/abs/1992PhRvA..46.3812B

**Fe thermal plasma (K$\alpha$)**

Tokamak measurement, includes both direct excitation and DR (F- to He-like Fe XVIII–XXV):
Beiersdorfer et al. 1993, ApJ 409, 846 — doi:10.1086/172715
http://adsabs.harvard.edu/abs/1993ApJ...409..846B

**Fe photo-exciation (K$\alpha$)**

In this measurement, K$\alpha$ transitions in L-shell ions of Fe were resonantly excited with a mono-energetic photon beam. Transitions not seen in this measurement are unlikely to be seen as absorption lines. (F- to He-like Fe XVIII–XXV):
Rudolph et al. 2013, PRL 111, 103002 — doi:10.1103/PhysRevLett.111.103002
http://adsabs.harvard.edu/abs/2013PhRvL.111j3002R

**Fe radiative and Auger rates (K$\alpha$)**

EBIT/photo-excitation measurement with ion extraction (Li- through C-like Fe XXI–XXIV):
Steinbrügge et al. 2015, PRA 91, 032502 — doi:10.1103/PhysRevA.91.032502
http://adsabs.harvard.edu/abs/2015PhRvA..91c2502S

**$Z$-trend for accuracy of theoretical He-w energies**

Measurements of He-w ($1s2p_{3/2}\,{}^1P_1 \rightarrow 1s^2\,{}^1S_0$) transition energy from multiple sources compared to various theoretical calculations as a function of $Z$ (C through Kr):
Beiersdorfer & Brown 2015, PRA 91, 032514 — doi:10.1103/PhysRevA.91.032514
`http://adsabs.harvard.edu/abs/2015PhRvA..91c2514B`

## Transition energies

See Table 1&2 for H- and He-like K-shell series and Table 3 for Li-like K$\alpha$ energy values.

**H-like ions: H–Ca (1 ≤ Z ≤ 20)** (Ly series and series limit / ionization potential):
Garcia & Mack 1965, JOSA 55, 654 — doi:10.1364/JOSA.55.000654
`http://adsabs.harvard.edu/abs/1965JOSA...55..654G`

**H-like ions: Sc–Ds (21 ≤ Z ≤ 110)** (Ly$\alpha$ and series limit / ionization potential):
Yerokhin & Shabaev 2015, JPCRD 44, 033103 — doi:10.1063/1.4927487
`https://ui.adsabs.harvard.edu/abs/2015JPCRD..44c3103Y`

**H-like ions: Sc–Ds (21 ≤ Z ≤ 110)** (Ly$\beta$+ ($n \geq 3$)):
The ground state of Erickson (1977) is not as accurate as that of Yerokhin & Shabaev (2015). The Erickson values are therefore adjusted to the ground state of Yerokhin & Shabaev (2015) (see reference for Ly$\alpha$).
Erickson 1977, JPCRD 6, 831 — doi:10.1063/1.555557
`http://adsabs.harvard.edu/abs/1977JPCRD...6..831E`

**He-like ions: He–B (2 ≤ Z ≤ 5)** (He$\alpha$ and series limit / ionization potential):
Yerokhin & Pachucki 2010, PRA 81, 022507 — doi:10.1103/PhysRevA.81.022507
`https://ui.adsabs.harvard.edu/abs/2010PhRvA..81b2507Y`

**He-like ions: C–U (6 ≤ Z ≤ 92)** (K-shell Rydberg series and series limit / ionization potential):
Yerokhin & Surzhykov 2019, JPCRD 48, 033104 — doi:10.1063/1.5121413
`https://ui.adsabs.harvard.edu/abs/2019JPCRD..48c3104Y`

**He-like ions: Np–Fm (93 ≤ Z ≤ 100)** (He$\alpha$ and series limit / ionization potential):
Artemyev et al. 2005, PRA 71, 062104 — doi:10.1103/PhysRevA.71.062104
`https://ui.adsabs.harvard.edu/abs/2005PhRvA..71f2104A`

**Li-like ions: C–Cl (6 ≤ Z ≤ 17)** (Li-$\alpha$):
Yerokhin et al. 2017, PRA 96, 042505 — doi:10.1103/PhysRevA.96.042505
`https://ui.adsabs.harvard.edu/abs/2017PhRvA..96d2505Y`
Erratum: doi:10.1103/PhysRevA.96.069901
`https://ui.adsabs.harvard.edu/abs/2017PhRvA..96f9901Y`

**Li-like ions: Ar–U (18 ≤ Z ≤ 92)** (Li-$\alpha$):
Yerokhin & Surzhykov 2018, JPCRD 47, 023105 — doi:10.1063/1.5034574
`https://ui.adsabs.harvard.edu/abs/2018JPCRD..47b3105Y`

## (Near-) neutral fluorescence lines: position and line shape

See Table 4 for parameters (positions, widths, amplitudes) used to create reference line shapes for neutral fluorescence lines.

**Neutral, empirical line shape** (Cr, Mn, Fe, Co, Ni, Cu K$\alpha$, K$\beta$):
Measurements in this paper are from neutral solid targets. Line shapes and energies for atomic samples can differ, but measurements for atomic lines are not available and theory is highly uncertain.
Caveat #1: for Mn K$\alpha$ see XRISM CALDB for updated reference made by F.S. Porter
Caveat #2: the exact line shape depends on excitation mechanism (collisions with electrons, protons, $\alpha$ particles; photoionization) and composition (gas, dust, atomic, molecular/minerals).

Hölzer et al. 1997, PRA 56, 4554 — doi:10.1103/PhysRevA.56.4554
http://adsabs.harvard.edu/abs/1997PhRvA..56.4554H

**Near-neutral / mildly ionized** (Fe II–IX)
The "neutral" fluorescent K$\alpha$ / K$\beta$ lines observed in XRISM spectra might not actually be neutral but a few times ionized, causing line shifts and differences in line shape. $\Delta E$ between neighboring charge states is larger for K$\beta$ than for K$\alpha$. For a theoretical model (no benchmark, so large uncertainties) for Fe II–IX see:
Palmeri et al. 2003, A&A 410, 359 — doi:10.1051/0004-6361:20031262
http://adsabs.harvard.edu/abs/2003A&A...410..359P

**Neutral line shape** (other elements; see Table 4 for details)

**F, Na, Al, Si, P, S, Cl, Ar, K, Ga, As, Se, Br, Rb, Y, Zr, Nb, Mo, Ag, Sn:**

– Line positions:
Bearden 1967, Rev. Modern Physics 39, 78 — doi:10.1103/RevModPhys.39.78
https://ui.adsabs.harvard.edu/abs/1967RvMP...39...78B
for F, Na, Al, Si, Cl, Ar, K, Ga, and Y from Bearden as listed in:
Zschornack, Handbook of X-ray Data, Springer 2007 — doi:10.1007/978-3-540-28619-6
https://link.springer.com/book/10.1007/978-3-540-28619-6

– Line positions (alternative):
Deslattes et al. 2003, Rev. Modern Physics 75, 35 — doi:10.1103/RevModPhys.75.35
https://ui.adsabs.harvard.edu/abs/2003RvMP...75...35D/

– 2-Lorentzian model (semi-empirical line widths):
Krause & Oliver 1979, JPCRD 8, 329 — doi:10.1063/1.555595
http://adsabs.harvard.edu/abs/1979JPCRD...8..329K

– Relative intensities (for F, Na): K$\alpha_1$/K$\alpha_2$ : 1.0/0.5

– Relative intensities (for Si, P, S, Cl, Ar, K, As, Se, Br, Rb, Zr, Mo, Ag, Sn):
Scofield 1974, PRA 9, 1041 — doi:10.1103/PhysRevA.9.1041
https://ui.adsabs.harvard.edu/abs/1974PhRvA...9.1041S

– Relative intensities (for Ga, Y, Nb):
Scofield 1974, ADNDT 14, 121 — doi:10.1016/S0092-640X(74)80019-7
https://ui.adsabs.harvard.edu/abs/1974ADNDT..14..121S

**Mg:**

– Line positions:
Schweppe et al. 1994, J. Electron Spectros. Relat. Phenomena 67, 463 —
doi:10.1016/0368-2048(93)02059-U
https://ui.adsabs.harvard.edu/abs/1994JESRP..67..463S

– 2-Lorentzian model (semi-empirical line widths):
Krause & Oliver 1979, JPCRD 8, 329 — doi:10.1063/1.555595
http://adsabs.harvard.edu/abs/1979JPCRD...8..329K

– Relative intensity: K$\alpha_1$/K$\alpha_2$ : 1.0/0.5

**Al (including satellites):** Note that the satellites are highly dependent on the details of the X-ray generator (see below), such that the here quoted values may not be a good match; and the K$\alpha_1$,K$\alpha_2$ positions from Bearden (1967) appear to be more the more accurate positions. We recommend using the 2-Lorentzian model for Al.

– Line positions:
Fischer & Baun 1965, J. Applied Physics 36, 534 — doi:10.1063/1.1714025
https://ui.adsabs.harvard.edu/abs/1965JAP....36..534F

– Line widths K$\alpha_1$, K$\alpha_2$:
Krause & Oliver 1979, JPCRD 8, 329 — doi:10.1063/1.555595
`http://adsabs.harvard.edu/abs/1979JPCRD...8..329K`

– Line widths K$\alpha_3$, K$\alpha_4$:
Nordfors 1955, Phys. Soc. A 68, 654 — doi:10.1088/0370-1298/68/7/416
`https://ui.adsabs.harvard.edu/abs/1955PPSA...68..654N`

– Line widths K$\alpha'$:
Wollman et al. 2000, NIMPRA 444, 145 — 10.1016/S0168-9002(99)01351-0
`https://ui.adsabs.harvard.edu/abs/2000NIMPA.444..145W`

**Ca, Sc, Ge:**
Ito et al. 2016, PRA 94, 042506 — doi:10.1103/PhysRevA.94.042506
`https://ui.adsabs.harvard.edu/abs/2016PhRvA..94d2506I`

**Ti, V:**
Chantler et al. 2006, PRA 73, 012508 — doi:10.1103/PhysRevA.73.012508
`https://ui.adsabs.harvard.edu/abs/2006PhRvA..73a2508C`

**Zn:**
Ito et al. 2015, JQSRT 151, 295 — doi:10.1016/j.jqsrt.2014.10.013
`https://ui.adsabs.harvard.edu/abs/2015JQSRT.151..295I`

**Additional background info on the line shapes:**

The complex line shape is caused by the fact that inner-shell transitions are often accompanied by the simultaneous movement of other bound electrons. Correspondingly, theoretical calculations find hundreds to thousands of transitions contributing to the overall line profile (e.g., Deutsch et al 2004). While the transition energies are somewhat constrained by theory, there is little constraint for their relative amplitudes, making the composition of a theoretical model a highly degenerate problem. Additionally, the exact line shapes can be highly variable as the relative positions and amplitudes of the many satellite lines have a strong dependence on source conditions, including variations due to chemical shifts due to the compound (molecule) being fluoresced; variations due to the excitation process (fluorescent emission induced by photons, electrons, or alpha-particles differ from each other); and variations due to the excitation energy (energy of the incident particles).

- Discussion of transitions contributing to K$\alpha$ fluorescent lines:
  Deutsch et al 2004, J. Res. Natl. Inst. Stand. Technol. 109, 75 — doi:10.6028/jres.109.006
  `https://pubmed.ncbi.nlm.nih.gov/27366598/`
- Reference energies of characteristic lines have uncertainties on the order of 1 eV:
  Bearden 1967, Rev. Modern Physics 39, 78 — doi:10.1103/RevModPhys.39.78
  `https://ui.adsabs.harvard.edu/abs/1967RvMP...39...78B`
- Line profile variations due to chemical shifts:
  Aberg et al 1970, JPhC 3, 1112 — doi:10.1088/0022-3719/3/5/024
  `https://ui.adsabs.harvard.edu/abs/1970JPhC....3.1112A`
  Deconninck & Van Den borek 1980, JPhC 13, 3329 — doi:10.1088/0022-3719/13/17/022
  `https://ui.adsabs.harvard.edu/abs/1980JPhC...13.3329D`
- Impact-energy dependence of relative intensity of satellite lines:
  Fischer & Baun 1965, J. Applied Physics 36, 534 — doi:10.1063/1.1714025
  `https://ui.adsabs.harvard.edu/abs/1965JAP....36..534F`

## Radiative branching ratios for K-shell transitions

These calculations are not great for level energies, but the transition rates appear fine. These papers include a table listing the energy levels and a table for transitions between energy levels. Excerpts of the tables are discussed in the paper. For full tables see the supplemental material linked on the journal's article webpage. K$\beta$ transitions are only included for ions where the $n = 2$ shell is filled in the ground state, i.e., for $N_e \geq 10$.
To obtain the **branching ratio** $\beta_r^j = A_r^{jk}/(A_r^j + A_a^j)$ of a given transition from upper level $j$ to lower level $k$, find the transition's radiative rate $A_r^{jk}$ from the transition table and the total radiative rate $A_r^j = \sum_i A_r^{ji}$ and auto-ionization (Auger) rate $A_a^j = \sum_l A_a^{jl}$ of the upper level $j$ in the energy level table.
To obtain the **fluorescence yield** $\omega_r^j = A_r^j/(A_r^j + A_a^j)$ of the excited level $j$, find the total radiative rate $A_r^j = \sum_i A_r^{ji}$ and auto-ionization (Auger) rate $A_a^j = \sum_l A_a^{jl}$ of the upper level $j$ in the energy level table.

**Fe** x–xxv ($2 \leq N_e \leq 17$)
Palmeri et al. 2003, A&A 403, 1175 — doi:10.1051/0004-6361:20030405
`http://adsabs.harvard.edu/abs/2003A&A...403.1175P`

**Fe** ii–ix ($18 \leq N_e \leq 25$)
Palmeri et al. 2003, A&A 410, 359 — doi:10.1051/0004-6361:20031262
`http://adsabs.harvard.edu/abs/2003A&A...410..359P`

**Ni** ii–xxvii ($2 \leq N_e \leq 27$) (K$\beta$ transitions only for $N_e > 9$)
Palmeri et al. 2008, ApJS 179, 542 — doi:10.1086/591965
`https://ui.adsabs.harvard.edu/abs/2008ApJS..179..542P`

**Ne, Mg, Si, S, Ar, Ca** ($1 \leq N_e \leq Z$) (K$\beta$ transitions only for $N_e > 9$)
Palmeri et al. 2008, ApJS 177, 408 — doi:10.1086/587804
`http://adsabs.harvard.edu/abs/2008ApJS..177..408P`

**F, Na, P, Cl, K, Sc, Ti, V, Cr, Mn, Co, Cu, Zn** ($2 \leq N_e \leq Z - 1$) (K$\beta$ transitions only for $N_e > 9$)
Palmeri et al. 2012, A&A 543, A44 — doi:10.1051/0004-6361/201219438
`https://ui.adsabs.harvard.edu/abs/2012A&A...543A..44P`

**Al** ($2 \leq N_e \leq 13$) (K$\beta$ transitions only for $N_e > 9$)
Palmeri et al. 2011, A&A 525, A59 — doi:10.1051/0004-6361/201014779
`https://ui.adsabs.harvard.edu/abs/2011A&A...525A..59P`

## Notation

**Letter designations for K$\alpha$ transitions in He- and Li-like ions:**
Gabriel 1972, MNRAS 160, 99 — doi:10.1093/mnras/160.1.99
`http://adsabs.harvard.edu/abs/1972MNRAS.160...99G`

**Letter designations for K$\alpha$ transitions in doubly-excited He-like ions, i.e., DR satellites to Ly$\alpha$, follow the convention of U. Safronova. For their definitions, see, e.g.:**
Bitter et al. 1984, PRA 29, 661 — doi:10.1103/PhysRevA.29.661
`http://adsabs.harvard.edu/abs/1984PhRvA..29..661B`

**Letter designations for L-shell transitions in Ne-like ions (e.g., 3C, 3D)**:
Parkinson 1973, A&A 24, 215 —
`http://adsabs.harvard.edu/abs/1973A%26A....24..215P`

## Historical values of $hc$

The value of the wavelength ↔ energy conversion constant $E\lambda = hc$ has historically changed as updated (and over time more accurate) values for its underlying constants (the Planck constant $h$, the speed of light $c$, and the elementary charge $e$) have become available. Published tables for transition and level energy / wavelength calculations and measurements can be decades old. The most reliable energy / wavelength values are in the "native" units the calculation or measurement was conducted in. For theory, those units are often Rydber (Ry), atomic units (a.u. = 2Ry), or Kaiser = $\mathrm{cm}^{-1}$. For convenience, these papers may also list their results converted to Å or eV. It is recommended to check which units the results were produced in and which values of physical constants were used in the conversion. The following table gives an overview of the evolution of the recommended values as a function of time. Only since the mid-1980s has the value of $hc$ stabilized to 7 significant digits, which corresponds to energy values being accurate to the meV level if converted from higher-precision wavelengths.

Table 0: Evolution of derived values for physical constants with time.

| $h[10^{-34}\,\mathrm{J\,s}]$ | $e[10^{-19}\,\mathrm{C}]$ | $hc/e[\mathrm{eV\,Å}]$ | source |
|---|---|---|---|
| 6.626 176(36) | 1.602 189 2(46) | 12398.521 | CODATA 1973[a] |
| 6.626 075 5(40) | 1.602 177 33(49) | 12398.4245 | CODATA 1986[b] |
| 6.626 068 76(52) | 1.602 176 462(63) | 12398.41857 | CODATA 1998[c] / ISIS |
| 6.626 068 96(33) | 1.602 176 87(40) | 12398.41579 | XDB 2009[d] with typo |
| 6.626 069 3(11) | 1.602 176 53(14) | 12398.41903 | CODATA 2002[e] |
| 6.626 068 96(33) | 1.602 176 487(40) | 12398.41875 | CODATA 2006[f] / XDB 2009[d] typo corr. |
| 6.626 069 57(29) | 1.602 176 565(35) | 12398.41929 | CODATA 2010[g] |
| 6.626 070 040(81) | 1.602 176 6208(98) | 12398.41974 | CODATA 2014[h] |
| 6.626 070 040 | 1.602 176 634 | 12398.41984 | CODATA 2018[i] |
| 6.626 070 040 | 1.602 176 634 | 12398.41984 | CODATA 2022[j] |

**Notes:**
The speed of light in vacuum is exact at $c = 299\,792\,458\,\mathrm{m\,s^{-1}}$ (after 1973).
The values of $h$ and $e$ have defined (exact) values as of CODATA 2018.

[a]CODATA 1973: Cohen 1976, ADNDT 18, 587 — doi:10.1016/0092-640X(76)90019-X
http://adsabs.harvard.edu/abs/1976ADNDT..18..587C

[b]CODATA 1986: Cohen & Taylor 1987, Rev. Mod. Phys. 59, 1121 — doi:10.1103/RevModPhys.59.1121
http://adsabs.harvard.edu/abs/1987RvMP...59.1121C

[c]CODATA 1998: Mohr & Taylor 2000, Rev. Mod. Phys. 72, 351 — doi:10.1103/RevModPhys.72.351
http://adsabs.harvard.edu/abs/2000RvMP...72..351M

[d]XDB 2009: X-ray Data Booklet; Thompson et al. 2009, https://xdb.lbl.gov/
It references CODATA 1998, but the quoted values match the numbers from CODATA 2006, with one exception: a digit of $e$ has been omitted. This has the consequence that the conversion constant derived from $h$, $c$, and $e$ does not yield the same number as the quoted value of $\hbar c$[MeV fm].

[e]CODATA 2002: Mohr & Taylor 2005, Rev. Mod. Phys. 77, 1 — doi:10.1103/RevModPhys.77.1
http://adsabs.harvard.edu/abs/2005RvMP...77....1M

[f]CODATA 2006: Mohr et al. 2008, Rev. Mod. Phys. 80, 633 — doi:10.1103/RevModPhys.80.633
http://adsabs.harvard.edu/abs/2008RvMP...80..633M

[g]CODATA 2010: Mohr et al. 2012, Rev. Mod. Phys. 84, 1527 — doi:10.1103/RevModPhys.84.1527
http://adsabs.harvard.edu/abs/2012RvMP...84.1527M

[h]CODATA 2014: Mohr et al. 2016, Rev. Mod. Phys. 88, 035009 — doi:10.1103/RevModPhys.88.035009
https://ui.adsabs.harvard.edu/abs/2016RvMP...88c5009M

[i]CODATA 2018: Tiesinga et al. 2021, Rev. Mod. Phys. 93, 025010 — doi:10.1103/RevModPhys.93.025010
https://ui.adsabs.harvard.edu/abs/2021RvMP...93b5010T

[j]CODATA 2022: Mohr et al. 2024, arXiv:2409.03787 — doi:10.48550/arXiv.2409.03787
https://ui.adsabs.harvard.edu/abs/2024arXiv240903787M

Table 1: Energies in eV for transitions to the ground state of H-like ions

| $Z$ | | Ly$\alpha_2$ $2p_{1/2}$ | Ly$\alpha_3$ $2s_{1/2}$ | Ly$\alpha_1$ $2p_{3/2}$ | Ly$\beta_2$ $3p_{1/2}$ | Ly$\beta_1$ $3p_{3/2}$ | Ly$\gamma_2$ $4p_{1/2}$ | Ly$\gamma_1$ $4p_{3/2}$ | Ly$\delta_2$ $5p_{1/2}$ | Ly$\delta_1$ $5p_{3/2}$ | Ly$\epsilon_2$ $6p_{1/2}$ | Ly$\epsilon_1$ $6p_{3/2}$ | Ly$\zeta_2$ $7p_{1/2}$ | Ly$\zeta_1$ $7p_{3/2}$ | Limit $n=\infty$ |
|---|---|---|---|---|---|---|---|---|---|---|---|---|---|---|---|
| 1 | H | 10.1988 | 10.1988 | 10.1989 | 12.0875 | 12.0875 | 12.7237 | 12.7485 | 13.0545 | 13.0545 | 13.2207 | 13.2207 | 13.3209 | 13.3209 | 13.5984 |
| 2 | He | 40.8130 | 40.8131 | 40.8138 | 48.3713 | 48.3715 | 51.0167 | 51.0167 | 52.2411 | 52.2411 | 52.9062 | 52.9062 | 53.3072 | 53.3072 | 54.4178 |
| 3 | Li | 91.8393 | 91.8396 | 91.8430 | 108.8481 | 108.8492 | 114.8010 | 114.8015 | 117.5563 | 117.5565 | 119.0530 | 119.0531 | 119.9554 | 119.9555 | 122.4544 |
| 4 | Be | 163.2846 | 163.2853 | 163.2962 | 193.5270 | 193.5304 | 204.1113 | 204.1128 | 209.0102 | 209.0109 | 211.6712 | 211.6716 | 213.2757 | 213.2760 | 217.7186 |
| 5 | B | 255.1592 | 255.1609 | 255.1876 | 302.4217 | 302.4301 | 318.9624 | 318.9659 | 326.6179 | 326.6197 | 330.7762 | 330.7773 | 333.2835 | 333.2841 | 340.2252 |
| 6 | C | 367.4740 | 367.4773 | 367.5329 | 435.5467 | 435.5642 | 459.3698 | 459.3772 | 470.3955 | 470.3992 | 476.3843 | 476.3864 | 479.9951 | 479.9965 | 489.9931 |
| 7 | N | 500.2466 | 500.2522 | 500.3557 | 592.9250 | 592.9574 | 625.3581 | 625.3717 | 640.3680 | 640.3750 | 648.5208 | 648.5249 | 653.4363 | 653.4389 | 667.0460 |
| 8 | O | 653.4937 | 653.5027 | 653.6799 | 774.5788 | 774.6339 | 816.9510 | 816.9743 | 836.5601 | 836.5720 | 847.2106 | 847.2175 | 853.6318 | 853.6362 | 871.4097 |
| 9 | F | 827.2374 | 827.2512 | 827.5360 | 980.5371 | 980.6256 | 1 034.1798 | 1 034.2172 | 1 059.0036 | 1 059.0228 | 1 072.4860 | 1 072.4970 | 1 080.6144 | 1 080.6215 | 1 103.1172 |
| 10 | Ne | 1 021.4979 | 1 021.5180 | 1 021.9531 | 1 210.8267 | 1 210.9617 | 1 277.0733 | 1 277.1302 | 1 307.7282 | 1 307.7573 | 1 324.3769 | 1 324.3938 | 1 334.4140 | 1 334.4247 | 1 362.1989 |
| 11 | Na | 1 236.3072 | 1 236.3353 | 1 236.9742 | 1 465.4894 | 1 465.6870 | 1 545.6759 | 1 545.7594 | 1 582.7793 | 1 582.8221 | 1 602.9295 | 1 602.9541 | 1 615.0771 | 1 615.0927 | 1 648.7017 |
| 12 | Mg | 1 471.6901 | 1 471.7284 | 1 472.6356 | 1 744.5580 | 1 744.8382 | 1 840.0232 | 1 840.1415 | 1 884.1936 | 1 884.2541 | 1 908.1805 | 1 908.2155 | 1 922.6407 | 1 922.6629 | 1 962.6632 |
| 13 | Al | 1 727.6852 | 1 727.7359 | 1 728.9887 | 2 048.0825 | 2 048.4687 | 2 160.1684 | 2 160.3313 | 2 212.0255 | 2 212.1088 | 2 240.1854 | 2 240.2337 | 2 257.1606 | 2 257.1911 | 2 304.1395 |
| 14 | Si | 2 004.3233 | 2 004.3892 | 2 006.0781 | 2 376.1038 | 2 376.6238 | 2 506.1556 | 2 506.3749 | 2 566.3205 | 2 566.4327 | 2 598.9900 | 2 599.0550 | 2 618.6827 | 2 618.7237 | 2 673.1772 |
| 15 | P | 2 301.6494 | 2 301.7330 | 2 303.9640 | 2 728.6802 | 2 729.3662 | 2 878.0467 | 2 878.3361 | 2 947.1418 | 2 947.2899 | 2 984.6582 | 2 984.7439 | 3 007.2717 | 3 007.3256 | 3 069.8416 |
| 16 | S | 2 619.7008 | 2 619.8056 | 2 622.7004 | 3 105.8611 | 3 106.7500 | 3 275.8950 | 3 276.2699 | 3 354.5442 | 3 354.7362 | 3 397.2456 | 3 397.3566 | 3 422.9831 | 3 423.0531 | 3 494.1889 |
| 17 | Cl | 2 958.5288 | 2 958.6581 | 2 962.3556 | 3 507.7127 | 3 508.8469 | 3 699.7711 | 3 700.2494 | 3 788.6002 | 3 788.8450 | 3 836.8253 | 3 836.9670 | 3 865.8907 | 3 865.9799 | 3 946.2937 |
| 18 | Ar | 3 318.1822 | 3 318.3396 | 3 322.9976 | 3 934.2988 | 3 935.7260 | 4 149.7433 | 4 150.3453 | 4 249.3796 | 4 249.6877 | 4 303.4682 | 4 303.6465 | 4 336.0658 | 4 336.1780 | 4 426.2279 |
| 19 | K | 3 698.7057 | 3 698.8951 | 3 704.6911 | 4 385.6785 | 4 384.9729 | 4 625.8750 | 4 626.6231 | 4 736.9476 | 4 737.3305 | 4 797.2399 | 4 797.4614 | 4 833.5742 | 4 833.7137 | 4 934.0582 |
| 20 | Ca | 4 100.1638 | 4 100.3894 | 4 107.5219 | 4 861.9351 | 4 864.1162 | 5 128.2549 | 5 129.1746 | 5 251.3950 | 5 251.8658 | 5 318.2326 | 5 318.5049 | 5 358.5089 | 5 358.6804 | 5 469.8785 |
| 21 | Sc | 4 522.5937 | 4 522.8633 | 4 531.5508 | 5 363.1255 | 5 365.7805 | 5 656.9461 | 5 658.0656 | 5 792.7872 | 5 793.3601 | 5 866.5125 | | 5 910.9366 | | 6 033.7564 |
| 22 | Ti | 4 966.0895 | 4 966.4061 | 4 976.8941 | 5 889.3595 | 5 892.5621 | 6 212.0628 | 6 213.4131 | 6 361.2403 | 6 361.9313 | 6 442.1967 | | 6 490.9747 | | 6 625.8102 |
| 23 | V | 5 430.7043 | 5 431.0732 | 5 443.6310 | 6 440.7115 | 6 444.5434 | 6 793.6860 | 6 795.3016 | 6 956.8378 | 6 957.6644 | 7 045.3697 | | 7 098.7084 | | 7 246.1262 |
| 24 | Cr | 5 916.5042 | 5 916.9315 | 5 931.8543 | 7 017.2688 | 7 021.8192 | 7 401.9090 | 7 403.8275 | 7 579.6754 | 7 580.6569 | 7 676.1284 | | 7 734.2352 | | 7 894.8029 |
| 25 | Mn | 6 423.5676 | 6 424.0594 | 6 441.6703 | 7 619.1326 | 7 624.4993 | 8 036.8402 | 8 039.1027 | 8 229.8640 | 8 231.0214 | 8 334.5852 | | 8 397.6681 | | 8 571.9544 |
| 26 | Fe | 6 951.9676 | 6 952.5304 | 6 973.1818 | 8 246.3989 | 8 252.6883 | 8 698.5820 | 8 701.2333 | 8 907.5088 | 8 908.8651 | 9 020.8461 | | 9 089.1138 | | 9 277.6886 |
| 27 | Co | 7 501.7886 | 7 502.4294 | 7 526.5042 | 8 899.1781 | 8 906.5053 | 9 387.2520 | 9 390.3408 | 9 612.7303 | 9 614.3104 | 9 735.0330 | | 9 808.6950 | | 10 012.1297 |
| 28 | Ni | 8 073.1106 | 8 073.8365 | 8 101.7498 | 11 973.2173 | 9 586.0644 | 10 102.9619 | 11 444.2486 | 10 345.6428 | 10 347.4736 | 10 477.2619 | | 10 556.5281 | | 10 775.3948 |
| 29 | Cu | 8 666.0278 | 8 666.8470 | 8 699.0468 | 10 281.7080 | 10 291.4989 | 10 845.8435 | 10 849.9700 | 11 106.3819 | 11 108.4925 | 11 247.6695 | | 11 332.7510 | | 11 567.6237 |
| 30 | Zn | 9 280.6268 | 9 281.5472 | 9 318.5175 | 11 011.6965 | 11 022.9320 | 11 616.0178 | 11 620.7524 | 11 895.0708 | 11 897.4926 | 12 046.3806 | | 12 137.4889 | | 12 388.9427 |
| 31 | Ga | 9 917.0103 | 9 918.0407 | 9 960.3018 | 11 767.6704 | 11 780.5089 | 12 413.6268 | 12 419.0364 | 12 711.8562 | 12 714.6229 | 12 873.5436 | | 12 970.8911 | | 13 239.5029 |
| 32 | Ge | 10 575.2742 | 10 576.4236 | 10 624.5342 | 12 549.7575 | 12 564.3665 | 13 238.8046 | 13 244.9608 | 13 556.8759 | 13 560.0238 | 13 729.2979 | | 13 833.0980 | | 14 119.4458 |
| 33 | As | 11 255.5242 | 11 256.8016 | 11 311.3612 | 13 358.0930 | 13 374.6535 | 14 091.7000 | 14 098.6773 | 14 430.2798 | 14 433.8476 | 14 613.7948 | | 14 724.2616 | | 15 028.9251 |
| 34 | Se | 11 957.8728 | 11 959.2882 | 12 020.9376 | 14 192.8261 | 14 211.5316 | 14 972.4685 | 14 980.3490 | 15 332.2298 | 15 336.2588 | 15 527.1980 | | 15 644.5468 | | 15 968.1075 |
| 35 | Br | 12 682.4267 | 12 683.9903 | 12 753.4140 | 15 054.0938 | 15 075.1513 | 15 881.2593 | 15 890.1304 | 16 262.8777 | 16 267.4131 | 16 469.6622 | | 16 594.1090 | | 16 937.1497 |
| 36 | Kr | 13 429.3148 | 13 431.0374 | 13 508.9654 | 15 942.0662 | 15 965.6951 | 16 818.2526 | 16 828.2049 | 17 222.4089 | 17 227.4968 | 17 441.3728 | | 17 573.1347 | | 17 936.2405 |
| 37 | Rb | 14 198.6516 | 14 200.5441 | 14 287.7540 | 16 856.8921 | 16 883.3256 | 17 783.6071 | 17 794.7409 | 18 210.9868 | 18 216.6773 | 18 442.4963 | | 18 581.7911 | | 18 965.5484 |
| 38 | Sr | 14 990.5687 | 14 992.6431 | 15 089.9615 | 17 798.7430 | 17 828.2314 | 18 777.5089 | 18 789.9284 | 19 228.7990 | 19 235.1458 | 19 473.2227 | | 19 620.2697 | | 20 025.2673 |
| 39 | Y | 15 805.1954 | 15 807.4638 | 15 915.7687 | 18 767.7857 | 18 800.5969 | 19 800.1364 | 19 813.9532 | 20 276.0312 | 20 283.0921 | 20 533.7398 | | 20 688.7593 | | 21 115.5877 |
| 40 | Zr | 16 642.6697 | 16 645.1451 | 16 765.3677 | 19 764.2076 | 19 800.6180 | | | | | | | | | 22 236.7129 |
| 41 | Nb | 17 503.1305 | 17 505.8272 | 17 638.9534 | 20 788.1821 | 20 828.4956 | | | | | | | | | 23 388.8499 |
| 42 | Mo | 18 386.7219 | 18 389.6555 | 18 536.7284 | 21 839.9108 | 21 884.4347 | | | | | | | | | 24 572.2125 |
| 43 | Tc | 19 293.6197 | 19 296.8022 | 19 458.9288 | 22 919.6149 | 22 968.6779 | | | | | | | | | 25 787.0471 |
| 44 | Ru | 20 223.9582 | 20 227.4070 | 20 405.7523 | 24 027.4687 | 24 081.4266 | | | | | | | | | 27 033.5647 |
| 45 | Rh | 21 177.9261 | 21 181.6549 | 21 377.4529 | 25 160.5644 | 25 219.8508 | | | | | | | | | 28 312.0310 |
| 46 | Pd | 22 155.6726 | 22 159.6995 | 22 374.2478 | 26 321.9529 | 26 386.9761 | | | | | | | | | 29 622.6774 |
| 47 | Ag | 23 157.3920 | 23 161.7328 | 23 396.4020 | 27 522.2310 | 27 593.1996 | | | | | | | | | 30 965.7805 |
| 48 | Cd | 24 183.2418 | 24 187.9176 | 24 444.1465 | 28 744.9788 | 28 822.4442 | | | | | | | | | 32 341.5869 |
| 49 | In | 25 233.4426 | 25 238.4699 | 25 517.7781 | 29 997.0558 | 30 081.5015 | | | | | | | | | 33 750.4047 |
| 50 | Sn | 26 308.1662 | 26 313.5664 | 26 617.5480 | 31 278.7164 | 31 370.6011 | | | | | | | | | 35 192.5008 |
| 51 | Sb | 27 407.6254 | 27 413.4188 | 27 743.7511 | 32 590.2184 | 32 690.0505 | | | | | | | | | 36 668.1827 |
| 52 | Te | 28 532.0046 | 28 538.2175 | 28 896.6576 | 34 303.7665 | 34 040.1389 | | | | | | | | | 38 177.7395 |
| 53 | I | 29 681.5849 | 29 688.2333 | 30 076.6372 | 35 303.8556 | 35 421.2314 | | | | | | | | | 39 721.5490 |
| 54 | Xe | 30 856.5321 | 30 863.6441 | 31 283.9483 | 36 706.5864 | 36 833.5834 | | | | | | | | | 41 299.8910 |
| 55 | Cs | 32 057.1148 | 32 064.7125 | 32 518.9553 | 38 177.5426 | 38 277.5855 | | | | | | | | | 42 913.1439 |
| 56 | Ba | 33 283.5434 | 33 291.6562 | 33 781.9681 | 39 605.4141 | 39 753.5380 | | | | | | | | | 44 561.6328 |
| 57 | La | 34 536.1019 | 34 544.7542 | 35 073.3747 | 41 102.1561 | 41 261.8477 | | | | | | | | | 46 245.7645 |
| 58 | Ce | 35 815.0238 | 35 824.2458 | 36 393.5160 | 42 630.8528 | 42 802.8561 | | | | | | | | | 47 965.8929 |
| 59 | Pr | 37 120.5968 | 37 130.4168 | 37 742.7918 | 44 192.0004 | 44 376.9724 | | | | | | | | | 49 722.4395 |
| 60 | Nd | 38 453.0709 | 38 463.5230 | 39 121.5685 | 45 785.8572 | 45 984.5666 | | | | | | | | | 51 515.7869 |
| 61 | Pm | 39 812.6826 | 39 823.8107 | 40 530.2033 | 47 412.6733 | 47 626.0005 | | | | | | | | | 53 346.3091 |
| 62 | Sm | 41 199.5825 | 41 211.4526 | 41 968.9716 | 49 072.7441 | 49 301.5569 | | | | | | | | | 55 214.3014 |
| 63 | Eu | 42 614.4976 | 42 627.1168 | 43 438.7345 | 50 766.9471 | 51 012.1134 | | | | | | | | | 57 120.6413 |
| 64 | Gd | 44 057.4766 | 44 070.8945 | 44 939.6751 | 52 495.3637 | 52 757.8630 | | | | | | | | | 59 065.5343 |

References:
$Z =$ 1– 20: Ly$n$ ($n \geq 2$) and series limit ($n = \infty$): J.D. Garcia & J.E. Mack (1965) JOSA 55, 654 – doi:10.1364/JOSA.55.000654
$Z =$ 21–110: Ly$\alpha$ ($n = 2$) and series limit ($n = \infty$): V.A. Yerokhin & V.M. Shabaev (2015) JPCRD 44, 033103 – doi:10.1063/1.4927487
$Z =$ 21–110: Ly$\beta$-$\zeta$ ($n =$ 3–7): G.W. Erickson (1977) JPCRD 6, 831 – doi:10.1063/1.555557 – corrected for the ground state of Yerokhin & Shabaev (2015)

Compiled by N. Hell, LLNL. hell1@llnl.gov . Cite the quoted references when using these values!

Table 1: Energies in eV for transitions to the ground state of H-like ions

| $Z$ | | Ly$\alpha_2$ $2p_{1/2}$ | Ly$\alpha_3$ $2s_{1/2}$ | Ly$\alpha_1$ $2p_{3/2}$ | Ly$\beta_2$ $3p_{1/2}$ | Ly$\beta_1$ $3p_{3/2}$ | Ly$\gamma_2$ $4p_{1/2}$ | Ly$\gamma_1$ $4p_{3/2}$ | Ly$\delta_2$ $5p_{1/2}$ | Ly$\delta_1$ $5p_{3/2}$ | Ly$\epsilon_2$ $6p_{1/2}$ | Ly$\epsilon_1$ $6p_{3/2}$ | Ly$\zeta_2$ $7p_{1/2}$ | Ly$\zeta_1$ $7p_{3/2}$ | Limit $n=\infty$ |
|---|---|---|---|---|---|---|---|---|---|---|---|---|---|---|---|
| 65 | Tb | 45 529.4180 | 45 543.6019 | 46 472.8384 | 54 259.1751 | 54 539.8753 | | | | | | | | | 61 050.0376 |
| 66 | Dy | 47 029.2239 | 47 044.3561 | 48 037.2632 | 56 057.2054 | 56 357.1232 | | | | | | | | | 63 073.2232 |
| 67 | Ho | 48 558.7596 | 48 574.8007 | 49 634.9806 | 57 891.3728 | 58 211.7480 | | | | | | | | | 65 137.1334 |
| 68 | Er | 50 117.5574 | 50 134.5982 | 51 265.6770 | 59 761.6302 | 60 103.3306 | | | | | | | | | 67 241.4729 |
| 69 | Tm | 51 706.6278 | 51 724.6656 | 52 930.5353 | 61 668.8136 | 62 033.0792 | | | | | | | | | 69 387.4499 |
| 70 | Yb | 53 325.3441 | 53 344.5336 | 54 629.0943 | 63 612.4864 | 64 000.6809 | | | | | | | | | 71 574.6276 |
| 71 | Lu | 54 974.8412 | 54 995.2320 | 56 362.6771 | 65 594.1072 | 66 007.3466 | | | | | | | | | 73 804.3409 |
| 72 | Hf | 56 655.9960 | 56 677.5733 | 58 132.3557 | 67 614.5398 | 68 054.1877 | | | | | | | | | 76 077.7012 |
| 73 | Ta | 58 368.5323 | 58 391.3940 | 59 938.0450 | 69 673.5824 | 70 141.1269 | | | | | | | | | 78 394.6310 |
| 74 | W | 60 112.9890 | 60 137.2262 | 61 780.4916 | 71 772.1334 | 72 268.8141 | | | | | | | | | 80 755.9044 |
| 75 | Re | 61 890.0327 | 61 915.7134 | 63 660.5806 | 73 910.8413 | 74 438.3941 | | | | | | | | | 83 162.4182 |
| 76 | Os | 63 699.6252 | 63 726.9033 | 65 578.4943 | 76 089.9565 | 76 649.8691 | | | | | | | | | 85 614.4226 |
| 77 | Ir | 65 543.2633 | 65 572.1482 | 67 535.9776 | 78 311.0625 | 78 904.9468 | | | | | | | | | 88 113.6252 |
| 78 | Pt | 67 420.4354 | 67 451.1080 | 69 532.7527 | 80 573.9688 | 81 203.5606 | | | | | | | | | 90 659.8354 |
| 79 | Au | 69 332.4801 | 69 365.0073 | 71 570.4286 | 82 879.9922 | 83 547.1512 | | | | | | | | | 93 254.6180 |
| 80 | Hg | 71 279.2548 | 71 313.8187 | 73 649.1248 | 85 229.3503 | 85 935.9362 | | | | | | | | | 95 898.1905 |
| 81 | Tl | 73 262.0492 | 73 298.7397 | 75 770.4245 | 87 623.4837 | 88 371.4804 | | | | | | | | | 98 592.1178 |
| 82 | Pb | 75 280.8320 | 75 319.8518 | 77 934.5857 | 90 062.5679 | 90 853.9590 | | | | | | | | | 101 336.6990 |
| 83 | Bi | 77 336.7135 | 77 378.2027 | 80 143.0387 | 92 547.9346 | 93 385.0759 | | | | | | | | | 104 133.3900 |
| 84 | Po | 79 430.4705 | 79 474.5679 | 82 396.8912 | 95 080.6995 | 95 965.6987 | | | | | | | | | 106 983.3066 |
| 85 | At | 81 562.4703 | 81 609.3964 | 84 696.8506 | 97 661.3630 | 98 596.5758 | | | | | | | | | 109 887.1968 |
| 86 | Rn | 83 729.5321 | 83 780.1303 | 87 040.0312 | 100 286.9983 | 101 275.0284 | | | | | | | | | 112 842.2582 |
| 87 | Fr | 85 940.1660 | 85 994.0623 | 89 435.4591 | 102 966.5852 | 104 009.7882 | | | | | | | | | 115 857.4702 |
| 88 | Ra | 88 190.6198 | 88 248.1991 | 91 879.7055 | 105 696.6390 | 106 797.7426 | | | | | | | | | 118 929.4725 |
| 89 | Ac | 90 485.1927 | 90 546.2959 | 94 377.5625 | 108 481.8386 | 109 643.4465 | | | | | | | | | 122 063.0677 |
| 90 | Th | 92 815.6031 | 92 881.5907 | 96 921.0121 | 111 350.4008 | 112 539.0374 | | | | | | | | | 125 250.2693 |
| 91 | Pa | 95 197.1844 | 95 267.0632 | 99 526.1581 | 114 208.0345 | 115 500.5698 | | | | | | | | | 128 507.1321 |
| 92 | U | 97 611.9514 | 97 687.9378 | 102 175.1036 | 117 147.5342 | 118 510.3685 | | | | | | | | | 131 816.1047 |
| 93 | Np | 100 084.1725 | 100 164.5685 | 104 893.1728 | 120 156.7391 | 121 593.4680 | | | | | | | | | 135 202.2216 |
| 94 | Pu | 102 588.6541 | 102 676.3156 | 107 655.1067 | 123 210.4107 | 124 724.6297 | | | | | | | | | 138 640.2442 |
| 95 | Am | 105 147.2727 | 105 241.4785 | 110 483.8800 | 126 331.8419 | 127 927.1466 | | | | | | | | | 142 153.4654 |
| 96 | Cm | 107 757.9195 | 107 858.5880 | 113 377.9429 | 129 519.2503 | 131 199.2362 | | | | | | | | | 145 740.1030 |
| 97 | Bk | 110 417.9549 | 110 525.7120 | 116 335.2246 | 132 769.3690 | 134 538.7475 | | | | | | | | | 149 398.1297 |
| 98 | Cf | 113 122.4946 | 113 239.0954 | 119 351.3714 | 136 079.5328 | 137 941.7755 | | | | | | | | | 153 123.6406 |
| 99 | Es | 115 881.5387 | 116 007.3097 | 122 437.3875 | 139 457.7385 | 141 419.1685 | | | | | | | | | 156 927.1121 |
| 100 | Fm | 118 692.8559 | 118 828.8584 | 125 591.7314 | 142 905.1052 | 144 968.2023 | | | | | | | | | 160 807.1836 |
| 101 | Md | 121 555.3032 | 121 703.0610 | 128 814.0221 | 146 419.0172 | 148 591.2204 | | | | | | | | | 164 763.7192 |
| 102 | No | 124 475.9013 | 124 636.1764 | 132 112.3959 | 150 007.6379 | 152 293.9065 | | | | | | | | | 168 804.8822 |
| 103 | Lr | 127 451.8746 | 127 626.2995 | 135 484.8663 | 153 670.0301 | 156 075.3236 | | | | | | | | | 172 928.4957 |
| 104 | Rf | 130 489.7537 | 130 679.1450 | 138 939.2157 | 157 411.7416 | 159 942.2591 | | | | | | | | | 177 142.5870 |
| 105 | Db | 133 586.1374 | 133 792.5357 | 142 472.9679 | 201 657.7150 | 163 893.3680 | | | | | | | | | 181 444.5711 |
| 106 | Sg | 136 740.7534 | 136 966.6079 | 146 086.9257 | | | | | | | | | | | 185 835.2643 |
| 107 | Bh | 139 965.5748 | 140 211.8378 | 149 794.8037 | | | | | | | | | | | 190 328.8166 |
| 108 | Hs | 143 246.7237 | 143 517.4770 | 153 583.4042 | | | | | | | | | | | 194 911.0780 |
| 109 | Mt | 146 603.1121 | 146 899.4441 | 157 473.8705 | | | | | | | | | | | 199 604.9060 |
| 110 | Ds | 150 019.0935 | 150 345.9023 | 161 451.3452 | | | | | | | | | | | 204 394.2461 |
| 111 | Rg | | | | | | | | | | | | | | |
| 112 | Cn | | | | | | | | | | | | | | |
| 113 | Nh | | | | | | | | | | | | | | |
| 114 | Fl | | | | | | | | | | | | | | |
| 115 | Mc | | | | | | | | | | | | | | |
| 116 | Lv | | | | | | | | | | | | | | |
| 117 | Ts | | | | | | | | | | | | | | |
| 118 | Og | | | | | | | | | | | | | | |

References:
$Z$ = 1– 20: Ly$n$ ($n \geq 2$) and series limit ($n = \infty$): J.D. Garcia & J.E. Mack (1965) JOSA 55, 654 – doi:10.1364/JOSA.55.000654
$Z$ = 21–110: Ly$\alpha$ ($n = 2$) and series limit ($n = \infty$): V.A. Yerokhin & V.M. Shabaev (2015) JPCRD 44, 033103 – doi:10.1063/1.4927487
$Z$ = 21–110: Ly$\beta$-$\zeta$ ($n$ = 3–7): G.W. Erickson (1977) JPCRD 6, 831 – doi:10.1063/1.555557 – corrected for the ground state of Yerokhin & Shabaev (2015)

Compiled by N. Hell, LLNL. hell1@llnl.gov . Cite the quoted references when using these values!

Table 2: Energies in eV for transitions to the ground state of He-like ions

| $Z$ | | Heα-z $2\,^3S_1$ | Heα-y $2\,^3P_1$ | Heα-x $2\,^3P_2$ | Heα-w $2\,^1P_1$ | Heβ$_2$ (y$_3$) $3\,^3P_1$ | Heβ$_1$ (w$_3$) $3\,^1P_1$ | Heγ$_2$ (y$_4$) $4\,^3P_1$ | Heγ$_1$ (w$_4$) $4\,^1P_1$ | Heδ$_2$ (y$_5$) $5\,^3P_1$ | Heδ$_1$ (w$_5$) $5\,^1P_1$ | Heϵ$_2$ (y$_6$) $6\,^3P_1$ | Heϵ$_1$ (w$_6$) $6\,^1P_1$ | Heζ$_2$ (y$_7$) $7\,^3P_1$ | Heζ$_1$ (w$_7$) $7\,^1P_1$ | Limit $n=\infty$ |
|---|---|---|---|---|---|---|---|---|---|---|---|---|---|---|---|---|
| 2 | He | 19.8196 | 20.9641 | 20.9641 | 21.2180 | | | | | | | | | | | 24.5874 |
| 3 | Li | 59.0207 | 61.2805 | 61.2807 | 62.2162 | | | | | | | | | | | 75.6401 |
| 4 | Be | 118.5928 | 121.9222 | 121.9241 | 123.6704 | | | | | | | | | | | 153.8962 |
| 5 | B | 198.5682 | 202.9545 | 202.9610 | 205.5652 | | | | | | | | | | | 259.3744 |
| 6 | C | 298.9622 | 304.4034 | 304.4202 | 307.9025 | 353.5316 | 354.5191 | 370.5121 | 370.9207 | 378.3222 | 378.5293 | 382.5483 | 382.6676 | 385.0899 | 385.1648 | 392.0906 |
| 7 | N | 419.7919 | 426.2923 | 426.3284 | 430.6957 | 496.6868 | 497.9194 | 521.0547 | 521.5635 | 532.2715 | 532.5293 | 538.3442 | 538.4925 | 541.9975 | 542.0906 | 552.0674 |
| 8 | O | 561.0723 | 568.6401 | 568.7084 | 573.9611 | 664.0937 | 665.5744 | 697.1757 | 697.7859 | 712.4132 | 712.7221 | 720.6658 | 720.8434 | 725.6318 | 725.7432 | 739.3270 |
| 9 | F | 722.8233 | 731.4702 | 731.5890 | 737.7216 | 855.7791 | 857.5108 | 898.9032 | 899.6160 | 918.7757 | 919.1364 | 929.5417 | 929.7491 | 936.0215 | 936.1515 | 953.8983 |
| 10 | Ne | 905.0621 | 914.8029 | 914.9960 | 922.0006 | 1 071.7672 | 1 073.7544 | 1 126.2629 | 1 127.0801 | 1 151.3854 | 1 151.7987 | 1 164.9986 | 1 165.2362 | 1 173.1933 | 1 173.3423 | 1 195.8082 |
| 11 | Na | 1 107.8177 | 1 118.6711 | 1 118.9696 | 1 126.8326 | 1 312.0968 | 1 314.3444 | 1 379.2954 | 1 380.2190 | 1 410.2834 | 1 410.7505 | 1 427.0782 | 1 427.3466 | 1 437.1892 | 1 437.3575 | 1 465.0992 |
| 12 | Mg | 1 331.1120 | 1 343.0990 | 1 343.5419 | 1 352.2485 | 1 576.7981 | 1 579.3133 | 1 658.0329 | 1 659.0661 | 1 695.5030 | 1 696.0253 | 1 715.8137 | 1 716.1140 | 1 728.0428 | 1 728.2309 | 1 761.8049 |
| 13 | Al | 1 574.9800 | 1 588.1256 | 1 588.7612 | 1 598.2915 | 1 865.9177 | 1 868.7093 | 1 962.5245 | 1 963.6709 | 2 007.0940 | 2 007.6736 | 2 031.2560 | 2 031.5890 | 2 045.8047 | 2 046.0135 | 2 085.9769 |
| 14 | Si | 1 839.4496 | 1 853.7806 | 1 854.6680 | 1 865.0016 | 2 179.4938 | 2 182.5736 | 2 292.8110 | 2 294.0756 | 2 345.0986 | 2 345.7380 | 2 373.4470 | 2 373.8145 | 2 390.5176 | 2 390.7480 | 2 437.6580 |
| 15 | P | 2 124.5620 | 2 140.1084 | 2 141.3190 | 2 152.4311 | 2 517.5809 | 2 520.9632 | 2 648.9499 | 2 650.3391 | 2 709.5753 | 2 710.2776 | 2 742.4465 | 2 742.8502 | 2 762.2413 | 2 762.4944 | 2 816.9087 |
| 16 | S | 2 430.3513 | 2 447.1441 | 2 448.7629 | 2 460.6293 | 2 880.2247 | 2 883.9284 | 3 030.9903 | 3 032.5120 | 3 100.5746 | 3 101.3440 | 3 138.3052 | 3 138.7474 | 3 161.0270 | 3 161.3044 | 3 223.7806 |
| 17 | Cl | 2 756.8650 | 2 774.9374 | 2 777.0653 | 2 789.6587 | 3 267.4875 | 3 271.5354 | 3 438.9988 | 3 440.6626 | 3 518.1645 | 3 519.0060 | 3 561.0919 | 3 561.5758 | 3 586.9440 | 3 587.2473 | 3 658.3437 |
| 18 | Ar | 3 104.1485 | 3 123.5347 | 3 126.2898 | 3 139.5824 | 3 679.4295 | 3 683.8499 | 3 873.0393 | 3 874.8575 | 3 962.4106 | 3 963.3305 | 4 010.8733 | 4 011.4022 | 4 040.0589 | 4 040.3906 | 4 120.6656 |
| 19 | K | 3 472.2419 | 3 492.9739 | 3 496.4939 | 3 510.4618 | 4 116.1035 | 4 120.9314 | 4 333.1693 | 4 335.1567 | 4 433.3722 | 4 434.3783 | 4 487.7090 | 4 488.2874 | 4 520.4321 | 4 520.7949 | 4 610.8071 |
| 20 | Ca | 3 861.2062 | 3 883.3172 | 3 887.7610 | 3 902.3780 | 4 577.5891 | 4 582.8655 | 4 819.4734 | 4 821.6476 | 4 931.1360 | 4 932.2370 | 4 991.6868 | 4 992.3200 | 5 028.1518 | 5 028.5492 | 5 128.8576 |
| 21 | Sc | 4 271.1001 | 4 294.6226 | 4 300.1724 | 4 315.4129 | 5 063.9621 | 5 069.7357 | 5 332.0333 | 5 334.4151 | 5 455.7858 | 5 456.9925 | 5 522.8917 | 5 523.5858 | 5 563.3037 | 5 563.7394 | 5 674.9036 |
| 22 | Ti | 4 701.9753 | 4 726.9381 | 4 733.8015 | 4 749.6449 | 5 575.2906 | 5 581.6186 | 5 870.9223 | 5 873.5361 | 6 007.3971 | 6 008.7220 | 6 081.3998 | 6 082.1624 | 6 125.9643 | 6 126.4428 | 6 249.0226 |
| 23 | V | 5 153.8971 | 5 180.3273 | 5 188.7387 | 5 205.1663 | 6 111.6591 | 6 118.6077 | 6 436.2316 | 6 439.1054 | 6 586.0632 | 6 587.5208 | 6 667.3062 | 6 668.1453 | 6 716.2293 | 6 716.7559 | 6 851.3112 |
| 24 | Cr | 5 626.9280 | 5 654.8495 | 5 665.0718 | 5 682.0688 | 6 673.1492 | 6 680.7940 | 7 028.0489 | 7 031.2150 | 7 191.8750 | 7 193.4817 | 7 280.7023 | 7 281.6275 | 7 334.1905 | 7 334.7713 | 7 481.8624 |
| 25 | Mn | 6 121.1431 | 6 150.5777 | 6 162.9042 | 6 180.4572 | 7 259.8578 | 7 268.2840 | 7 646.4787 | 7 649.9731 | 7 824.9387 | 7 826.7133 | 7 921.6960 | 7 922.7182 | 7 979.9568 | 7 980.5986 | 8 140.7864 |
| 26 | Fe | 6 636.6121 | 6 667.5780 | 6 682.3333 | 6 700.4340 | 7 871.8758 | 7 881.1791 | 8 291.6182 | 8 295.4816 | 8 485.3554 | 8 487.3183 | 8 590.3893 | 8 591.5204 | 8 653.6306 | 8 654.3408 | 8 828.1864 |
| 27 | Co | 7 173.4158 | 7 205.9292 | 7 223.4711 | 7 242.1126 | 8 509.3078 | 8 519.5943 | 8 963.5804 | 8 967.8575 | 9 173.2401 | 9 175.4145 | 9 286.8989 | 9 288.1521 | 9 355.3294 | 9 356.1166 | 9 544.1817 |
| 28 | Ni | 7 731.6304 | 7 765.7043 | 7 786.4241 | 7 805.6048 | 9 172.2534 | 9 183.6401 | 9 662.4721 | 9 667.2123 | 9 888.7029 | 9 891.1140 | 10 011.3359 | 10 012.7259 | 10 085.1652 | 10 086.0384 | 10 288.8848 |
| 29 | Cu | 8 311.3476 | 8 346.9936 | 8 371.3187 | 8 391.0355 | 9 860.8314 | 9 873.4458 | 10 388.4204 | 10 393.6777 | 10 631.8746 | 10 634.5500 | 10 763.8330 | 10 765.3757 | 10 843.2710 | 10 844.2403 | 11 062.4309 |
| 30 | Zn | 8 912.6484 | 8 949.8754 | 8 978.2691 | 8 998.5252 | 10 575.1481 | 10 589.1297 | 11 141.5401 | 11 147.3732 | 11 402.8722 | 11 405.8419 | 11 544.5082 | 11 546.2210 | 11 629.7661 | 11 630.8423 | 11 864.9402 |
| 31 | Ga | 9 535.6315 | 9 574.4479 | 9 607.4117 | 9 628.2090 | 11 315.3324 | 11 330.8317 | 11 921.9686 | 11 928.4411 | 12 201.8369 | 12 205.1335 | 12 353.5047 | 12 355.4064 | 12 444.7942 | 12 445.9893 | 12 696.5581 |
| 32 | Ge | 10 180.3895 | 10 220.8019 | 10 258.8761 | 10 280.2198 | 12 081.5050 | 12 098.6848 | 12 729.8357 | 12 737.0160 | 13 028.9025 | 13 032.5608 | 13 190.9576 | 13 193.0684 | 13 288.4914 | 13 289.8179 | 13 557.4218 |
| 33 | As | 10 847.0241 | 10 889.0382 | 10 932.8032 | 10 954.7010 | 12 873.7980 | 12 892.8337 | 13 565.2829 | 13 573.2449 | 13 884.2139 | 13 888.2711 | 14 057.0130 | 14 059.3546 | 14 161.0040 | 14 162.4762 | 14 447.6799 |
| 34 | Se | 11 535.6440 | 11 579.2653 | 11 629.3440 | 11 651.8019 | 13 692.3559 | 13 713.4339 | 14 428.4634 | 14 437.2853 | 14 767.9268 | 14 772.4235 | 14 951.8295 | 14 954.4241 | 15 062.4928 | 15 064.1241 | 15 367.4935 |
| 35 | Br | 12 246.3513 | 12 291.5842 | 12 348.6422 | 12 371.6703 | 14 537.3089 | 14 560.6318 | 15 319.5193 | 15 329.2868 | 15 680.1885 | 15 685.1682 | 15 875.5554 | 15 878.4289 | 15 993.1059 | 15 994.9127 | 16 317.0137 |
| 36 | Kr | 12 979.2711 | 13 026.1211 | 13 090.8701 | 13 114.4749 | 15 408.8230 | 15 434.6030 | 16 238.6260 | 16 249.4276 | 16 621.1773 | 16 626.6863 | 16 828.3716 | 16 831.5513 | 16 953.0256 | 16 955.0243 | 17 296.4238 |
| 37 | Rb | 13 734.5130 | 13 782.9847 | 13 856.1819 | 13 880.3741 | 16 307.0423 | 16 335.5081 | 17 185.9374 | 17 197.8696 | 17 591.0523 | 17 597.1382 | 17 810.4373 | 17 813.9503 | 17 942.4139 | 17 944.6221 | 18 305.8863 |
| 38 | Sr | 14 512.2051 | 14 562.3036 | 14 644.7556 | 14 669.5439 | 17 232.1328 | 17 263.5266 | 18 161.6316 | 18 174.7978 | 18 589.9959 | 18 596.7117 | 18 821.9390 | 18 825.8166 | 18 961.4545 | 18 963.8913 | 19 345.5900 |
| 39 | Y | 15 312.4722 | 15 364.2026 | 15 456.7659 | 15 482.1605 | 18 184.2592 | 18 218.8394 | 19 165.8855 | 19 180.3919 | 19 618.1891 | 19 625.5892 | 19 863.0589 | 19 867.3311 | 20 010.3323 | 20 013.0167 | 20 415.7191 |
| 40 | Zr | 16 135.4482 | 16 188.8159 | 16 292.3995 | 16 318.4104 | 19 163.5956 | 19 201.6358 | 20 198.8855 | 20 214.8477 | 20 675.8222 | 20 683.9652 | 20 933.9888 | 20 938.6896 | 21 089.2420 | 21 092.1972 | 21 516.4713 |
| 41 | Nb | 16 981.2669 | 17 036.2773 | 17 151.8448 | 17 178.4813 | 20 170.3203 | 20 212.1089 | 21 260.8207 | 21 278.3598 | 21 763.0912 | 21 772.0397 | 22 034.9275 | 22 040.0936 | 22 198.3809 | 22 201.6285 | 22 648.0460 |
| 42 | Mo | 17 850.0706 | 17 906.7289 | 18 035.3006 | 18 062.5720 | 21 204.6210 | 21 250.4641 | 22 351.8914 | 22 371.1364 | 22 880.1989 | 22 890.0181 | 23 166.0817 | 23 171.7498 | 23 337.9583 | 23 341.5217 | 23 810.6532 |
| 43 | Tc | 18 742.0256 | 18 800.3404 | 18 942.9947 | 18 970.9127 | 22 266.7087 | 22 316.9300 | 23 472.3241 | 23 493.4115 | 24 027.3792 | 24 038.1386 | 24 327.6827 | 24 333.8937 | 24 508.2085 | 24 512.1120 | 25 004.5313 |
| 44 | Ru | 19 657.2675 | 19 717.2447 | 19 875.1221 | 19 903.6953 | 23 356.7641 | 23 411.7094 | 24 622.3154 | 24 645.3893 | 25 204.8322 | 25 216.6039 | 25 519.9387 | 25 526.7348 | 25 709.3381 | 25 713.6090 | 26 229.8880 |
| 45 | Rh | 20 595.9743 | 20 657.6234 | 20 831.9271 | 20 861.1667 | 24 475.0229 | 24 535.0498 | 25 802.1099 | 25 827.3199 | 26 412.8096 | 26 425.6724 | 26 743.1028 | 26 750.5274 | 26 941.6003 | 26 946.2657 | 27 486.9791 |
| 46 | Pd | 21 558.2956 | 21 621.6238 | 21 813.6236 | 21 843.5390 | 25 621.6807 | 25 687.1717 | 27 011.9213 | 27 039.4280 | 27 651.5291 | 27 665.5633 | 27 997.3926 | 28 005.4934 | 28 205.2195 | 28 210.3094 | 28 776.0314 |
| 47 | Ag | 22 544.4174 | 22 609.4350 | 22 820.4687 | 22 851.0705 | 26 796.9680 | 26 868.3350 | 28 252.0068 | 28 281.9801 | 28 921.2559 | 28 936.5473 | 29 283.0789 | 29 291.9035 | 29 500.4625 | 29 506.0082 | 30 097.3143 |
| 48 | Cd | 23 554.4992 | 23 621.2133 | 23 852.6905 | 23 883.9878 | 28 001.1314 | 28 078.7750 | 29 522.5976 | 29 555.2131 | 30 222.2241 | 30 238.8639 | 30 600.3957 | 30 609.9986 | 30 827.5686 | 30 833.6027 | 31 451.0699 |
| 49 | In | 24 588.7455 | 24 657.1683 | 24 910.5718 | 24 942.5753 | 29 234.4032 | 29 318.7802 | 30 823.9726 | 30 859.4168 | 31 554.7235 | 31 572.8041 | 31 949.6383 | 31 960.0725 | 32 186.8304 | 32 193.3870 | 32 837.5934 |
| 50 | Sn | 25 647.3298 | 25 717.4734 | 25 994.3609 | 26 027.0802 | 30 497.0251 | 30 588.6036 | 32 156.3809 | 32 194.8510 | 32 919.0058 | 32 938.6293 | 33 331.0610 | 33 342.3837 | 33 578.5091 | 33 585.6235 | 34 257.1475 |
| 51 | Sb | 26 730.4556 | 26 802.2965 | 27 104.3424 | 27 137.7888 | 31 789.2625 | 31 888.5351 | 33 520.1067 | 33 561.8081 | 34 315.3649 | 34 336.6360 | 34 744.7086 | 34 757.2329 | 35 002.8987 | 35 010.6076 | 35 710.0301 |
| 52 | Te | 27 838.3091 | 27 911.9035 | 28 240.7834 | 28 274.9653 | 33 111.3658 | 33 218.8481 | 34 915.4195 | 34 960.5673 | 35 744.0756 | 35 767.1019 | 36 191.6131 | 36 204.8963 | 36 460.2793 | 36 468.6250 | 37 196.5242 |
| 53 | I | 28 971.1527 | 29 046.5011 | 29 404.0404 | 29 438.9710 | 34 463.6697 | 34 579.9045 | 36 342.6690 | 36 391.4930 | 37 205.4957 | 37 230.3942 | 37 671.3833 | 37 685.7468 | 37 951.0170 | 37 960.0376 | 38 716.9985 |
| 54 | Xe | 30 129.1588 | 30 206.2687 | 30 594.3691 | 30 630.0569 | 35 846.4040 | 35 971.9642 | 37 802.1122 | 37 854.8492 | 38 699.8893 | 38 726.7810 | 39 184.5363 | 39 200.0468 | 39 475.3784 | 39 485.1201 | 40 271.7265 |
| 55 | Cs | 31 312.5845 | 31 391.4702 | 31 812.1245 | 31 848.5810 | 37 259.9158 | 37 395.3959 | 39 294.1098 | 39 351.0115 | 40 227.6263 | 40 256.6391 | 40 731.4465 | 40 748.1788 | 41 033.7378 | 41 044.2455 | 41 861.0769 |
| 56 | Ba | 32 521.6392 | 32 602.3106 | 33 057.6093 | 33 094.8441 | 38 704.4731 | 38 850.5002 | 40 818.9556 | 40 880.2819 | 41 789.0075 | 41 820.2720 | 42 312.4158 | 42 330.4461 | 42 626.4039 | 42 637.7252 | 43 485.3694 |
| 57 | La | 33 756.5939 | 33 839.0667 | 34 331.2003 | 34 369.2252 | 40 180.4323 | 40 337.6651 | 42 377.0265 | 42 443.0549 | 43 384.4179 | 43 418.0756 | 43 927.8374 | 43 947.2446 | 44 253.7672 | 44 265.9525 | 45 144.9959 |
| 58 | Ce | 35 017.6812 | 35 101.9676 | 35 633.2315 | 35 672.0572 | 41 688.1022 | 41 857.2292 | 43 968.6572 | 44 039.6721 | 45 014.2017 | 45 050.3969 | 45 578.0556 | 45 598.9240 | 45 916.1761 | 45 929.2784 | 46 840.3075 |
| 59 | Pr | 36 305.1802 | 36 391.2954 | 36 964.0942 | 37 003.7317 | 43 227.8476 | 43 409.5898 | 45 594.2341 | 45 670.5390 | 46 678.7521 | 46 717.6372 | 47 263.4717 | 47 285.8885 | 47 614.0319 | 47 628.1043 | 48 571.7122 |
| 60 | Nd | 37 619.3404 | 37 707.2967 | 38 324.1480 | 38 364.6084 | 44 800.0018 | 44 995.1143 | 47 254.1124 | 47 336.0214 | 48 378.4379 | 48 420.1733 | 48 984.4545 | 49 008.5121 | 49 347.7088 | 49 362.8098 | 50 339.5870 |
| 61 | Pm | 38 960.3927 | 39 050.1940 | 39 713.7327 | 39 755.0268 | 46 404.8845 | 46 614.1564 | 48 948.6416 | 49 036.4868 | 50 113.6155 | 50 158.3687 | 50 741.3685 | 50 767.1622 | 51 117.5728 | 51 133.7622 | 52 144.2897 |
| 62 | Sm | 40 328.5166 | 40 420.1473 | 41 133.1305 | 41 175.2690 | 48 042.7351 | 48 266.9964 | 50 678.0885 | 50 772.2140 | 51 884.5611 | 51 932.5075 | 52 534.4914 | 52 562.1205 | 52 923.9027 | 52 941.2418 | 53 986.1209 |
| 63 | Eu | 41 724.3539 | 41 817.8574 | 42 583.1713 | 42 626.1669 | 49 714.3591 | 49 954.4724 | 52 443.2802 | 52 544.0440 | 53 692.1196 | 53 743.4402 | 54 364.6762 | 54 394.2414 | 54 767.5544 | 54 786.0989 | 55 865.9270 |
| 64 | Gd | 43 147.9961 | 43 243.3772 | 44 064.0402 | 44 107.9037 | 51 419.9144 | 51 676.7899 | 54 244.4155 | 54 352.1998 | 55 536.4801 | 55 591.3655 | 56 232.1120 | 56 263.7316 | 56 648.7183 | 56 668.5554 | 57 783.9134 |

References:

$Z$ = 2–5: Kα ($n$ = 2) and series limit ($n=\infty$): V.A. Yerokhin & K. Pachucki (2010) PRA 81, 022507 – doi:10.1103/PhysRevA.81.022507
$Z$ = 6–92: Kα–ζ ($n$ = 2–7) and series limit ($n=\infty$): V.A. Yerokhin & A. Surzhykov (2019) JPCRD 48, 033104 – doi:10.1063/1.5121413
$Z$ = 93-100: Kα ($n$ = 2) and series limit ($n=\infty$): A.N. Artemyev, V.M. Shabaev, V.A. Yerokhin, G. Plunien & G. Soff (2005) PRA 71, 062104 – doi:10.1103/PhysRevA.71.062104

Compiled by N. Hell, LLNL. hell1@llnl.gov . Cite the quoted references when using these values!

Table 2: Energies in eV for transitions to the ground state of He-like ions

| $Z$ | | He$\alpha$-z $2\ ^3S_1$ | He$\alpha$-y $2\ ^3P_1$ | He$\alpha$-x $2\ ^3P_2$ | He$\alpha$-w $2\ ^1P_1$ | He$\beta_2$ ($y_3$) $3\ ^3P_1$ | He$\beta_1$ ($w_3$) $3\ ^1P_1$ | He$\gamma_2$ ($y_4$) $4\ ^3P_1$ | He$\gamma_1$ ($w_4$) $4\ ^1P_1$ | He$\delta_2$ ($y_5$) $5\ ^3P_1$ | He$\delta_1$ ($w_5$) $5\ ^1P_1$ | He$\epsilon_2$ ($y_6$) $6\ ^3P_1$ | He$\epsilon_1$ ($w_6$) $6\ ^1P_1$ | He$\zeta_2$ ($y_7$) $7\ ^3P_1$ | He$\zeta_1$ ($w_7$) $7\ ^1P_1$ | Limit $n = \infty$ |
|---|---|---|---|---|---|---|---|---|---|---|---|---|---|---|---|---|
| 65 | Tb | 44 600.2374 | 44 697.5832 | 45 576.7548 | 45 621.4987 | 53 160.3683 | 53 434.9447 | 56 082.4766 | 56 197.6760 | 57 418.6509 | 57 477.3050 | 58 137.8206 | 58 171.6035 | 58 568.4272 | 58 589.6113 | 59 741.1155 |
| 66 | Dy | 46 080.2173 | 46 179.4066 | 47 120.3806 | 47 166.0172 | 54 934.7547 | 55 228.0390 | 57 956.5383 | 58 079.5610 | 59 337.7339 | 59 400.3609 | 60 080.9041 | 60 116.9591 | 60 525.7558 | 60 548.3821 | 61 736.6203 |
| 67 | Ho | 47 589.5322 | 47 690.6623 | 48 696.8930 | 48 743.4339 | 56 745.0466 | 57 058.0456 | 59 868.6008 | 59 999.8686 | 61 295.7156 | 61 362.5195 | 62 063.3352 | 62 101.8121 | 62 522.7314 | 62 546.8543 | 63 772.4211 |
| 68 | Er | 49 127.8599 | 49 230.9050 | 50 305.9967 | 50 353.4553 | 58 590.8630 | 58 924.6923 | 61 818.3239 | 61 958.2992 | 63 292.2694 | 63 363.5088 | 64 084.8283 | 64 125.8358 | 64 559.0275 | 64 584.7422 | 65 848.2445 |
| 69 | Tm | 50 696.0847 | 50 801.1103 | 51 948.8353 | 51 997.2233 | 60 473.3195 | 60 829.0811 | 63 806.8367 | 63 956.0095 | 65 328.5655 | 65 404.4580 | 66 146.5397 | 66 190.2276 | 66 635.8141 | 66 663.2159 | 67 965.2536 |
| 70 | Yb | 52 293.7686 | 52 400.6934 | 53 624.9857 | 53 674.3164 | 62 391.9671 | 62 770.8720 | 65 833.7585 | 65 992.5913 | 67 404.1956 | 67 484.9862 | 68 248.0751 | 68 294.5658 | 68 752.7103 | 68 781.8673 | 70 123.0485 |
| 71 | Lu | 53 921.8571 | 54 030.6975 | 55 335.6719 | 55 385.9580 | 64 347.9896 | 64 751.2487 | 67 900.2864 | 68 069.2963 | 69 520.3979 | 69 606.3587 | 70 390.6588 | 70 440.1291 | 70 910.9405 | 70 941.9615 | 72 322.8714 |
| 72 | Hf | 55 581.2388 | 55 692.0880 | 57 082.0517 | 57 133.3068 | 66 342.4617 | 66 771.3812 | 70 007.5497 | 70 187.2809 | 71 678.3152 | 71 769.7047 | 72 575.4610 | 72 628.0470 | 73 111.6750 | 73 144.6416 | 74 565.9059 |
| 73 | Ta | 57 271.6767 | 57 384.5123 | 58 863.9564 | 58 916.1941 | 68 375.1931 | 68 831.1199 | 72 155.3853 | 72 346.3956 | 73 877.7980 | 73 974.9018 | 74 802.3321 | 74 858.1834 | 75 354.7640 | 75 389.7715 | 76 852.0025 |
| 74 | W | 58 993.7071 | 59 108.5181 | 60 682.1369 | 60 735.3706 | 70 446.8912 | 70 931.2266 | 74 344.5413 | 74 547.4158 | 76 119.5944 | 76 222.6984 | 77 072.0338 | 77 131.3410 | 77 640.9830 | 77 678.1402 | 79 181.9366 |
| 75 | Re | 60 747.9395 | 60 864.7419 | 62 537.4415 | 62 591.6845 | 72 558.3179 | 73 072.5450 | 76 575.8478 | 76 791.1851 | 78 404.5618 | 78 513.9787 | 79 385.4234 | 79 448.3361 | 79 971.1757 | 80 010.5914 | 81 556.5791 |
| 76 | Os | 62 534.4718 | 62 653.2018 | 64 430.1035 | 64 485.3702 | 74 709.6909 | 75 255.3336 | 78 849.5359 | 79 077.9755 | 80 732.9448 | 80 849.0014 | 81 742.7458 | 81 809.4817 | 82 345.6140 | 82 387.4107 | 83 976.1883 |
| 77 | Ir | 64 354.5747 | 64 475.3100 | 66 361.7743 | 66 418.0795 | 76 902.5477 | 77 481.2250 | 81 167.2113 | 81 409.4335 | 83 106.3900 | 83 229.4127 | 84 145.6744 | 84 216.3832 | 84 765.9444 | 84 810.2309 | 86 442.4379 |
| 78 | Pt | 66 207.9545 | 66 330.6192 | 68 332.2373 | 68 389.5951 | 79 136.6433 | 79 750.0287 | 83 528.6835 | 83 785.3685 | 85 524.6795 | 85 655.0220 | 86 593.9781 | 86 668.8911 | 87 231.9763 | 87 278.8751 | 88 955.1373 |
| 79 | Au | 68 095.7240 | 68 220.3425 | 70 342.9652 | 70 401.3903 | 81 413.4336 | 82 063.2823 | 85 935.4354 | 86 207.3179 | 87 989.3508 | 88 127.3805 | 89 089.2350 | 89 168.5426 | 89 745.2471 | 89 794.9079 | 91 515.7832 |
| 80 | Hg | 70 017.9935 | 70 144.4951 | 72 394.2301 | 72 453.7370 | 83 733.0138 | 84 421.1625 | 88 387.6302 | 88 675.4587 | 90 500.5808 | 90 646.6651 | 91 631.5812 | 91 715.5011 | 92 305.9610 | 92 358.4926 | 94 124.6748 |
| 81 | Tl | 71 975.7615 | 72 104.1573 | 74 487.3952 | 74 547.9992 | 86 096.8125 | 86 825.1660 | 90 886.7512 | 91 191.3282 | 93 059.8797 | 93 214.4131 | 94 222.5678 | 94 311.3313 | 94 915.6146 | 94 971.1666 | 96 783.2001 |
| 82 | Pb | 73 969.2568 | 74 099.4677 | 76 622.8826 | 76 684.5989 | 88 505.0066 | 89 275.5786 | 93 433.0566 | 93 755.1986 | 95 667.4924 | 95 830.9104 | 96 862.4532 | 96 956.2917 | 97 574.4936 | 97 633.2157 | 99 491.8078 |
| 83 | Bi | 75 999.3745 | 76 131.3696 | 78 801.9442 | 78 864.7889 | 90 958.9566 | 91 773.8424 | 96 027.9208 | 96 368.4985 | 98 324.8611 | 98 497.5718 | 99 552.6797 | 99 651.8380 | 100 284.0537 | 100 346.0957 | 102 251.7771 |
| 84 | Po | 78 066.9406 | 78 200.7119 | 81 025.7508 | 81 089.7388 | 93 459.6013 | 94 320.9914 | 98 672.4050 | 99 032.3028 | 101 033.0335 | 101 215.4995 | 102 294.3221 | 102 399.0451 | 103 045.3428 | 103 110.8678 | 105 064.3052 |
| 85 | At | 80 172.3127 | 80 307.7743 | 83 294.9125 | 83 360.0597 | 96 007.4851 | 96 917.7059 | 101 367.0397 | 101 747.3188 | 103 792.6627 | 103 985.3193 | 105 087.9247 | 105 198.5389 | 105 858.9593 | 105 928.2123 | 107 930.0315 |
| 86 | Rn | 82 313.1509 | 82 449.5962 | 85 606.7581 | 85 673.0796 | 98 599.9819 | 99 561.3193 | 104 109.3624 | 104 510.7984 | 106 601.0818 | 106 804.4869 | 107 931.1200 | 108 047.8160 | 108 722.4951 | 108 795.4760 | 110 846.3166 |
| 87 | Fr | 84 496.4496 | 84 634.3937 | 87 970.0006 | 88 037.5132 | 101 245.4864 | 102 260.5120 | 106 907.8086 | 107 331.5307 | 109 466.9714 | 109 681.6012 | 110 832.3163 | 110 955.4206 | 111 644.3993 | 111 721.3667 | 113 821.8953 |
| 88 | Ra | 86 719.3097 | 86 858.5242 | 90 381.3138 | 90 450.0333 | 103 940.7062 | 105 012.0321 | 109 759.1265 | 110 206.2095 | 112 387.0389 | 112 613.4376 | 113 788.3436 | 113 918.1692 | 114 621.5019 | 114 702.6462 | 116 853.4750 |
| 89 | Ac | 88 985.3569 | 89 126.1622 | 92 845.3495 | 92 915.2921 | 106 690.1311 | 107 820.4921 | 112 667.9284 | 113 139.5017 | 115 365.9374 | 115 604.7173 | 116 803.9231 | 116 940.7963 | 117 658.4966 | 117 743.9404 | 119 945.7361 |
| 90 | Th | 91 288.1695 | 91 429.3789 | 95 354.4381 | 95 425.6232 | 109 486.1010 | 110 678.3815 | 115 626.6360 | 116 123.8697 | 118 396.0749 | 118 647.7122 | 119 871.2722 | 120 015.4789 | 120 747.6557 | 120 837.7662 | 123 091.0458 |
| 91 | Pa | 93 640.0278 | 93 782.7842 | 97 923.9238 | 97 996.3618 | 112 343.5687 | 113 600.7347 | 118 650.2156 | 119 174.4430 | 121 492.5810 | 121 757.7559 | 123 005.6701 | 123 157.6457 | 123 904.3261 | 123 999.1578 | 126 304.7650 |
| 92 | U | 96 026.9344 | 96 169.2404 | 100 536.7745 | 100 610.4889 | 115 245.9215 | 116 571.2113 | 121 722.3539 | 122 274.6498 | 124 638.8975 | 124 918.2088 | 126 190.7220 | 126 350.7114 | 127 111.8411 | 127 211.7749 | 129 569.9136 |
| 93 | Np | 98 468.4100 | 98 612.2600 | 103 217.5100 | 103 292.5100 | | | | | | | | | | | 132 911.0000 |
| 94 | Pu | 100 945.8900 | 101 088.8300 | 105 943.1200 | 106 019.4200 | | | | | | | | | | | 136 305.1000 |
| 95 | Am | 103 472.9700 | 103 615.3200 | 108 730.8700 | 108 808.5000 | | | | | | | | | | | 139 769.5000 |
| 96 | Cm | 106 054.7600 | 106 197.6400 | 111 587.4500 | 111 666.4100 | | | | | | | | | | | 143 310.9000 |
| 97 | Bk | 108 681.0400 | 108 823.0800 | 114 500.4000 | 114 580.7100 | | | | | | | | | | | 146 916.9000 |
| 98 | Cf | 111 355.6000 | 111 495.9200 | 117 474.7900 | 117 556.4800 | | | | | | | | | | | 150 592.6000 |
| 99 | Es | 114 082.9000 | 114 221.2100 | 120 516.4900 | 120 599.5700 | | | | | | | | | | | 154 343.9000 |
| 100 | Fm | 116 862.8000 | 116 998.4800 | 123 625.7700 | 123 710.2400 | | | | | | | | | | | 158 171.1000 |
| 101 | Md | | | | | | | | | | | | | | | |
| 102 | No | | | | | | | | | | | | | | | |
| 103 | Lr | | | | | | | | | | | | | | | |
| 104 | Rf | | | | | | | | | | | | | | | |
| 105 | Db | | | | | | | | | | | | | | | |
| 106 | Sg | | | | | | | | | | | | | | | |
| 107 | Bh | | | | | | | | | | | | | | | |
| 108 | Hs | | | | | | | | | | | | | | | |
| 109 | Mt | | | | | | | | | | | | | | | |
| 110 | Ds | | | | | | | | | | | | | | | |
| 111 | Rg | | | | | | | | | | | | | | | |
| 112 | Cn | | | | | | | | | | | | | | | |
| 113 | Nh | | | | | | | | | | | | | | | |
| 114 | Fl | | | | | | | | | | | | | | | |
| 115 | Mc | | | | | | | | | | | | | | | |
| 116 | Lv | | | | | | | | | | | | | | | |
| 117 | Ts | | | | | | | | | | | | | | | |
| 118 | Og | | | | | | | | | | | | | | | |

References:
$Z = 2$–5: K$\alpha$ ($n = 2$) and series limit ($n = \infty$): V.A. Yerokhin & K. Pachucki (2010) PRA 81, 022507 – doi:10.1103/PhysRevA.81.022507
$Z = 6$–92: K$\alpha$–$\zeta$ ($n = 2$–7) and series limit ($n = \infty$): V.A. Yerokhin & A. Surzhykov (2019) JPCRD 48, 033104 – doi:10.1063/1.5121413
$Z = 93$-100: K$\alpha$ ($n = 2$) and series limit ($n = \infty$): A.N. Artemyev, V.M. Shabaev, V.A. Yerokhin, G. Plunien & G. Soff (2005) PRA 71, 062104 – doi:10.1103/PhysRevA.71.062104

Compiled by N. Hell, LLNL. hell1@llnl.gov . Cite the quoted references when using these values!

Table 3: Energies in eV for transitions to the $n = 2$ shell of Li-like ions

| Key | Upper | Lower | 6 | 7 | 8 | 9 | 10 | 11 | 12 | 13 | 12 | 15 | 16 | 17 |
|---|---|---|---|---|---|---|---|---|---|---|---|---|---|---|
| | | | C | N | O | F | Ne | Na | Mg | Al | Si | P | S | Cl |
| a | $1s(^2S)2p^2(^3P)\ ^2P_{3/2}$ | $1s^22p\ ^2P^o_{3/2}$ | 299.7386 | 420.7732 | 562.2994 | 724.3312 | 906.8866 | 1 109.9973 | 1 333.6920 | 1 578.0126 | 1 842.9981 | 2 128.7034 | 2 435.1782 | 2 762.4863 |
| b | $1s(^2S)2p^2(^3P)\ ^2P_{3/2}$ | $1s^22p\ ^2P^o_{1/2}$ | 299.7517 | 420.8051 | 562.3652 | 724.4519 | 907.0914 | 1 110.3229 | 1 334.1856 | 1 578.7326 | 1 844.0148 | 2 130.0986 | 2 437.0488 | 2 764.9450 |
| c | $1s(^2S)2p^2(^3P)\ ^2P_{1/2}$ | $1s^22p\ ^2P^o_{3/2}$ | 299.7205 | 420.7345 | 562.2269 | 724.2059 | 906.6825 | 1 109.6807 | 1 333.2189 | 1 577.3272 | 1 842.0299 | 2 127.3651 | 2 433.3630 | 2 760.0646 |
| d | $1s(^2S)2p^2(^3P)\ ^2P_{1/2}$ | $1s^22p\ ^2P^o_{1/2}$ | 299.7336 | 420.7664 | 562.2925 | 724.3265 | 906.8873 | 1 110.0063 | 1 333.7124 | 1 578.0470 | 1 843.0466 | 2 128.7600 | 2 435.2336 | 2 762.5233 |
| e | $1s(^2S)2p^2(^3P)\ ^4P_{5/2}$ | $1s^22p\ ^2P^o_{3/2}$ | 295.3134 | 415.0377 | 555.2351 | 715.9280 | 897.1368 | 1 098.8975 | 1 321.2362 | 1 564.1954 | 1 827.8074 | 2 112.1234 | 2 417.1815 | 2 743.0357 |
| f | $1s(^2S)2p^2(^3P)\ ^4P_{3/2}$ | $1s^22p\ ^2P^o_{3/2}$ | 295.3085 | 415.0236 | 555.2043 | 715.8694 | 897.0354 | 1 098.7341 | 1 320.9869 | 1 563.8312 | 1 827.2948 | 2 111.4231 | 2 416.2492 | 2 741.8243 |
| g | $1s(^2S)2p^2(^3P)\ ^4P_{3/2}$ | $1s^22p\ ^2P^o_{1/2}$ | 295.3216 | 415.0554 | 555.2700 | 715.9900 | 897.2402 | 1 099.0597 | 1 321.4805 | 1 564.5511 | 1 828.3114 | 2 112.8180 | 2 418.1201 | 2 744.2828 |
| h | $1s(^2S)2p^2(^3P)\ ^4P_{1/2}$ | $1s^22p\ ^2P^o_{3/2}$ | 295.2990 | 415.0039 | 555.1678 | 715.8067 | 896.9343 | 1 098.5791 | 1 320.7584 | 1 563.5050 | 1 826.8403 | 2 110.8044 | 2 415.4226 | 2 740.7358 |
| i | $1s(^2S)2p^2(^3P)\ ^4P_{1/2}$ | $1s^22p\ ^2P^o_{1/2}$ | 295.3121 | 415.0358 | 555.2334 | 715.9274 | 897.1391 | 1 098.9048 | 1 321.2520 | 1 564.2248 | 1 827.8570 | 2 112.1993 | 2 417.2931 | 2 743.1947 |
| **j** | $\mathbf{1s(^2S)2p^2(^1D)\ ^2D_{5/2}}$ | $\mathbf{1s^22p\ ^2P^o_{3/2}}$ | **298.4757** | **419.1988** | **560.3920** | **722.0776** | **904.2714** | **1 107.0057** | **1 330.3043** | **1 574.2074** | **1 838.7474** | **2 123.9754** | **2 429.9330** | **2 756.6795** |
| **k** | $\mathbf{1s(^2S)2p^2(^1D)\ ^2D_{3/2}}$ | $\mathbf{1s^22p\ ^2P^o_{1/2}}$ | **298.5022** | **419.2514** | **560.4943** | **722.2522** | **904.5505** | **1 107.4270** | **1 330.9150** | **1 575.0619** | **1 839.9071** | **2 125.5058** | **2 431.9053** | **2 759.1672** |
| l | $1s(^2S)2p^2(^1D)\ ^2D_{3/2}$ | $1s^22p\ ^2P^o_{3/2}$ | 298.4891 | 419.2195 | 560.4287 | 722.1315 | 904.3453 | 1 107.1013 | 1 330.4215 | 1 574.3421 | 1 838.8906 | 2 124.1108 | 2 430.0345 | 2 756.7083 |
| m | $1s(^2S)2p^2(^1S)\ ^2S_{1/2}$ | $1s^22p\ ^2P^o_{3/2}$ | 304.6123 | 426.6678 | 569.1944 | 732.2085 | 915.7275 | 1 119.7863 | 1 344.4073 | 1 589.6323 | 1 855.4942 | 2 142.0446 | 2 449.3280 | 2 777.4033 |
| n | $1s(^2S)2p^2(^1S)\ ^2S_{1/2}$ | $1s^22p\ ^2P^o_{1/2}$ | 304.6250 | 426.6997 | 569.2600 | 732.3291 | 915.9325 | 1 120.1121 | 1 344.9009 | 1 590.3523 | 1 856.5108 | 2 143.4400 | 2 451.1986 | 2 779.8618 |
| o | $1s2s^2\ ^2S_{1/2}$ | $1s^22p\ ^2P^o_{3/2}$ | 283.5889 | 400.9108 | 538.6851 | 696.9237 | 875.6255 | 1 074.8121 | 1 294.4927 | 1 534.6885 | 1 795.4210 | 2 076.7131 | 2 378.5855 | 2 701.0707 |
| p | $1s2s^2\ ^2S_{1/2}$ | $1s^22p\ ^2P^o_{1/2}$ | 283.6018 | 400.9427 | 538.7507 | 697.0440 | 875.8309 | 1 075.1374 | 1 294.9864 | 1 535.4084 | 1 796.4379 | 2 078.1083 | 2 380.4561 | 2 703.5292 |
| **q** | $\mathbf{1s(^2S)2s2p(^3P^o)\ ^2P^o_{3/2}}$ | $\mathbf{1s^22s\ ^2S_{1/2}}$ | **299.9698** | **421.2842** | **563.0793** | **725.3753** | **908.1936** | **1 111.5705** | **1 335.5332** | **1 580.1259** | **1 845.3860** | **2 131.3648** | **2 438.1088** | **2 765.6771** |
| **r** | $\mathbf{1s(^2S)2s2p(^3P^o)\ ^2P^o_{1/2}}$ | $\mathbf{1s^22s\ ^2S_{1/2}}$ | **299.9576** | **421.2592** | **563.0332** | **725.2968** | **908.0687** | **1 111.3794** | **1 335.2532** | **1 579.7283** | **1 844.8358** | **2 130.6198** | **2 437.1182** | **2 764.3803** |
| s | $1s(^2S)2s2p(^1P^o)\ ^2P^o_{3/2}$ | $1s^22s\ ^2S_{1/2}$ | 303.4172 | 425.3326 | 567.7216 | 730.6090 | 914.0168 | 1 117.9826 | 1 342.5339 | 1 587.7196 | 1 853.5752 | 2 140.1559 | 2 447.5100 | 2 775.7008 |
| **t** | $\mathbf{1s(^2S)2s2p(^1P^o)\ ^2P^o_{1/2}}$ | $\mathbf{1s^22s\ ^2S_{1/2}}$ | **303.4182** | **425.3317** | **567.7158** | **730.5943** | **913.9866** | **1 117.9285** | **1 342.4454** | **1 587.5832** | **1 853.3746** | **2 139.8726** | **2 447.1221** | **2 775.1845** |
| **u** | $\mathbf{1s(^2S)2s2p(^3P^o)\ ^4P^o_{3/2}}$ | $\mathbf{1s^22s\ ^2S_{1/2}}$ | **294.0892** | **413.8623** | **554.0963** | **714.8132** | **896.0331** | **1 097.7898** | **1 320.1074** | **1 563.0263** | **1 826.5779** | **2 110.8084** | **2 415.7558** | **2 741.4733** |
| **v** | $\mathbf{1s(^2S)2s2p(^3P^o)\ ^4P^o_{1/2}}$ | $\mathbf{1s^22s\ ^2S_{1/2}}$ | **294.0886** | **413.8580** | **554.0842** | **714.7873** | **895.9852** | **1 097.7095** | **1 319.9811** | **1 562.8379** | **1 826.3078** | **2 110.4347** | **2 415.2523** | **2 740.8109** |

| Key | Upper | Lower | 18 | 19 | 20 | 21 | 22 | 23 | 24 | 25 | 26 | 27 | 28 | 29 |
|---|---|---|---|---|---|---|---|---|---|---|---|---|---|---|
| | | | Ar | K | Ca | Sc | Ti | V | Cr | Mn | Fe | Co | Ni | Cu |
| a | $1s(^2S)2p^2(^3P)\ ^2P_{3/2}$ | $1s^22p\ ^2P^o_{3/2}$ | 3 110.7086 | 3 479.8812 | 3 870.0967 | 4 281.4411 | 4 713.9897 | 5 167.8373 | 5 643.0722 | 6 139.7982 | 6 658.1159 | 7 198.1363 | 7 759.9704 | 8 343.7396 |
| b | $1s(^2S)2p^2(^3P)\ ^2P_{3/2}$ | $1s^22p\ ^2P^o_{1/2}$ | 3 113.8798 | 3 483.9147 | 3 875.1613 | 4 287.7220 | 4 721.6948 | 5 177.1977 | 5 654.3438 | 6 153.2626 | 6 674.0811 | 7 216.9399 | 7 781.9778 | 8 369.3496 |
| c | $1s(^2S)2p^2(^3P)\ ^2P_{1/2}$ | $1s^22p\ ^2P^o_{3/2}$ | 3 107.5260 | 3 475.7556 | 3 864.8200 | 4 274.7703 | 4 705.6577 | 5 157.5447 | 5 630.4914 | 6 124.5719 | 6 639.8620 | 7 176.4451 | 7 734.4084 | 8 313.8485 |
| d | $1s(^2S)2p^2(^3P)\ ^2P_{1/2}$ | $1s^22p\ ^2P^o_{1/2}$ | 3 110.6961 | 3 479.7893 | 3 869.8829 | 4 281.0508 | 4 713.3625 | 5 166.9052 | 5 641.7631 | 6 138.0364 | 6 655.8273 | 7 195.2485 | 7 756.4159 | 8 339.4586 |
| e | $1s(^2S)2p^2(^3P)\ ^4P_{5/2}$ | $1s^22p\ ^2P^o_{3/2}$ | 3 089.7541 | 3 457.3461 | 3 845.8916 | 4 255.4501 | 4 686.0671 | 5 137.8034 | 5 610.7137 | 6 104.8635 | 6 620.3110 | 7 157.1257 | 7 715.3719 | 8 295.1328 |
| f | $1s(^2S)2p^2(^3P)\ ^4P_{3/2}$ | $1s^22p\ ^2P^o_{3/2}$ | 3 088.2102 | 3 455.4161 | 3 843.5202 | 4 252.5835 | 4 682.6549 | 5 133.8020 | 5 606.0863 | 6 099.5839 | 6 614.3645 | 7 150.5090 | 7 708.0935 | 8 287.2108 |
| g | $1s(^2S)2p^2(^3P)\ ^4P_{3/2}$ | $1s^22p\ ^2P^o_{1/2}$ | 3 091.3803 | 3 459.4500 | 3 848.5848 | 4 258.8648 | 4 690.3590 | 5 143.1626 | 5 617.3578 | 6 113.0483 | 6 630.3295 | 7 169.3125 | 7 730.1012 | 8 312.8206 |
| h | $1s(^2S)2p^2(^3P)\ ^4P_{1/2}$ | $1s^22p\ ^2P^o_{3/2}$ | 3 086.7971 | 3 453.6027 | 3 841.2161 | 4 249.6799 | 4 679.0233 | 5 129.2902 | 5 600.5156 | 6 092.7441 | 6 606.0100 | 7 140.3564 | 7 695.8160 | 8 272.4380 |
| i | $1s(^2S)2p^2(^3P)\ ^4P_{1/2}$ | $1s^22p\ ^2P^o_{1/2}$ | 3 089.9674 | 3 457.6363 | 3 846.2793 | 4 255.9599 | 4 686.7279 | 5 138.6507 | 5 611.7874 | 6 106.2087 | 6 621.9754 | 7 159.1594 | 7 717.8237 | 8 298.0481 |
| **j** | $\mathbf{1s(^2S)2p^2(^1D)\ ^2D_{5/2}}$ | $\mathbf{1s^22p\ ^2P^o_{3/2}}$ | **3 104.2909** | **3 472.7891** | **3 862.2700** | **4 272.8169** | **4 704.5025** | **5 157.4255** | **5 631.6802** | **6 127.3802** | **6 644.6350** | **7 183.5673** | **7 744.2956** | **8 326.9557** |
| **k** | $\mathbf{1s(^2S)2p^2(^1D)\ ^2D_{3/2}}$ | $\mathbf{1s^22p\ ^2P^o_{1/2}}$ | **3 107.3656** | **3 476.5314** | **3 866.7546** | **4 278.1112** | **4 710.6727** | **5 164.5256** | **5 639.7598** | **6 136.4706** | **6 654.7581** | **7 194.7312** | **7 756.4974** | **8 340.1789** |
| l | $1s(^2S)2p^2(^1D)\ ^2D_{3/2}$ | $1s^22p\ ^2P^o_{3/2}$ | 3 104.1945 | 3 472.4974 | 3 861.6902 | 4 271.8320 | 4 702.9678 | 5 155.1652 | 5 628.4870 | 6 123.0060 | 6 638.7926 | 7 175.9280 | 7 734.4894 | 8 314.5689 |
| m | $1s(^2S)2p^2(^1S)\ ^2S_{1/2}$ | $1s^22p\ ^2P^o_{3/2}$ | 3 126.3493 | 3 496.1890 | 3 887.0212 | 4 298.9272 | 4 731.9811 | 5 186.2780 | 5 661.9141 | 6 158.9911 | 6 677.6178 | 7 217.9045 | 7 779.9708 | 8 363.9437 |
| n | $1s(^2S)2p^2(^1S)\ ^2S_{1/2}$ | $1s^22p\ ^2P^o_{1/2}$ | 3 129.5200 | 3 500.2240 | 3 892.0850 | 4 305.2072 | 4 739.6854 | 5 195.6386 | 5 673.1864 | 6 172.4557 | 6 693.5833 | 7 236.7078 | 7 801.9781 | 8 389.5538 |
| o | $1s2s^2\ ^2S_{1/2}$ | $1s^22p\ ^2P^o_{3/2}$ | 3 044.2110 | 3 408.0019 | 3 792.4970 | 4 197.7329 | 4 623.7399 | 5 070.5613 | 5 538.2329 | 6 026.8012 | 6 536.3034 | 7 066.8015 | 7 618.3370 | 8 190.9619 |
| p | $1s2s^2\ ^2S_{1/2}$ | $1s^22p\ ^2P^o_{1/2}$ | 3 047.3812 | 3 412.0357 | 3 797.5605 | 4 204.0128 | 4 631.4450 | 5 079.9226 | 5 549.5045 | 6 040.2664 | 6 552.2691 | 7 085.6049 | 7 640.3447 | 8 216.5723 |
| **q** | $\mathbf{1s(^2S)2s2p(^3P^o)\ ^2P^o_{3/2}}$ | $\mathbf{1s^22s\ ^2S_{1/2}}$ | **3 114.1426** | **3 483.5369** | **3 873.9469** | **4 285.4486** | **4 718.1138** | **5 172.0294** | **5 647.2813** | **6 143.9674** | **6 662.1877** | **7 202.0521** | **7 763.6707** | **8 347.1667** |
| **r** | $\mathbf{1s(^2S)2s2p(^3P^o)\ ^2P^o_{1/2}}$ | $\mathbf{1s^22s\ ^2S_{1/2}}$ | **3 112.4680** | **3 481.3977** | **3 871.2411** | **4 282.0592** | **4 713.9001** | **5 166.8307** | **5 640.9068** | **6 136.2037** | **6 652.7873** | **7 190.7362** | **7 750.1245** | **8 331.0413** |
| s | $1s(^2S)2s2p(^1P^o)\ ^2P^o_{3/2}$ | $1s^22s\ ^2S_{1/2}$ | 3 124.8049 | 3 494.8547 | 3 885.9467 | 4 298.1627 | 4 731.5766 | 5 186.2801 | 5 662.3666 | 6 159.9400 | 6 679.0999 | 7 219.9612 | 7 782.6334 | 8 367.2451 |
| **t** | $\mathbf{1s(^2S)2s2p(^1P^o)\ ^2P^o_{1/2}}$ | $\mathbf{1s^22s\ ^2S_{1/2}}$ | **3 124.1333** | **3 494.0008** | **3 884.8788** | **4 296.8503** | **4 729.9899** | **5 184.3891** | **5 660.1383** | **6 157.3434** | **6 676.1073** | **7 216.5446** | **7 778.7657** | **8 362.8988** |
| **u** | $\mathbf{1s(^2S)2s2p(^3P^o)\ ^4P^o_{3/2}}$ | $\mathbf{1s^22s\ ^2S_{1/2}}$ | **3 088.0226** | **3 455.4209** | **3 843.7466** | **4 253.0590** | **4 683.4119** | **5 134.8717** | **5 607.4998** | **6 101.3699** | **6 616.5480** | **7 153.1134** | **7 711.1353** | **8 290.7020** |
| **v** | $\mathbf{1s(^2S)2s2p(^3P^o)\ ^4P^o_{1/2}}$ | $\mathbf{1s^22s\ ^2S_{1/2}}$ | **3 087.1706** | **3 454.3446** | **3 842.4089** | **4 251.4242** | **4 681.4384** | **5 132.5228** | **5 604.7335** | **6 098.1505** | **6 612.8404** | **7 148.8862** | **7 706.3606** | **8 285.3572** |

References:
$Z = 6$–17: K$\alpha$ ($n = 2$): V.A. Yerokhin, A. Surzhykov & A. Müller (2017) PRA 96, 042505 – doi:10.1103/PhysRevA.96.042505
$Z = 18$–92: K$\alpha$ ($n = 2$): V.A. Yerokhin & A. Surzhykov (2018) JPCRD 47, 023105 – doi:10.1063/1.5034574

Compiled by N. Hell, LLNL. hell1@llnl.gov . Cite the quoted references when using these values!

Table 3: Energies in eV for transitions to the $n = 2$ shell of Li-like ions

| Key | Upper | Lower | 30 Zn | 31 Ga | 32 Ge | 33 As | 34 Se | 35 Br | 36 Kr | 37 Rb | 38 Sr | 39 Y | 40 Zr | 41 Nb |
|---|---|---|---|---|---|---|---|---|---|---|---|---|---|---|
| a | $1s(^2S)2p^2(^3P)\,^2P_{3/2}$ | $1s^22p\,^2P^o_{3/2}$ | 8 949.5646 | 9 577.5772 | 10 227.9109 | 10 900.7073 | 11 596.1170 | 12 314.2850 | 13 055.3796 | 13 819.5613 | 14 607.0031 | 15 417.8846 | 16 252.3896 | 17 110.7044 |
| b | $1s(^2S)2p^2(^3P)\,^2P_{3/2}$ | $1s^22p\,^2P^o_{1/2}$ | 8 979.2076 | 9 611.7199 | 10 267.0535 | 10 945.3888 | 11 646.9165 | 12 371.8217 | 13 120.3167 | 13 892.6041 | 14 688.9083 | 15 509.4514 | 16 354.4695 | 17 224.2068 |
| c | $1s(^2S)2p^2(^3P)\,^2P_{1/2}$ | $1s^22p\,^2P^o_{3/2}$ | 8 914.8592 | 9 537.5476 | 10 182.0176 | 10 848.3824 | 11 536.7525 | 12 247.2423 | 12 979.9760 | 13 735.0702 | 14 512.6537 | 15 312.8522 | 16 135.8025 | 16 981.6314 |
| d | $1s(^2S)2p^2(^3P)\,^2P_{1/2}$ | $1s^22p\,^2P^o_{1/2}$ | 8 944.5021 | 9 571.6901 | 10 221.1604 | 10 893.0638 | 11 587.5519 | 12 304.7780 | 13 044.9127 | 13 808.1136 | 14 594.5578 | 15 404.4179 | 16 237.8837 | 17 095.1334 |
| e | $1s(^2S)2p^2(^3P)\,^4P_{5/2}$ | $1s^22p\,^2P^o_{3/2}$ | 8 896.4785 | 9 519.5046 | 10 164.2972 | 10 830.9572 | 11 519.5914 | 12 230.3009 | 12 963.2128 | 13 718.4381 | 14 496.1048 | 15 296.3406 | 16 119.2800 | 16 965.0546 |
| f | $1s(^2S)2p^2(^3P)\,^4P_{3/2}$ | $1s^22p\,^2P^o_{3/2}$ | 8 887.9403 | 9 510.3829 | 10 154.6287 | 10 820.7823 | 11 508.9506 | 12 219.2369 | 12 951.7673 | 13 706.6496 | 14 484.0169 | 15 283.9897 | 16 106.7052 | 16 952.2966 |
| g | $1s(^2S)2p^2(^3P)\,^4P_{3/2}$ | $1s^22p\,^2P^o_{1/2}$ | 8 917.5831 | 9 544.5248 | 10 193.7720 | 10 865.4637 | 11 559.7496 | 12 276.7730 | 13 016.7028 | 13 779.6936 | 14 565.9206 | 15 375.5566 | 16 208.7862 | 17 065.7978 |
| h | $1s(^2S)2p^2(^3P)\,^4P_{1/2}$ | $1s^22p\,^2P^o_{3/2}$ | 8 870.2535 | 9 489.3155 | 10 129.6655 | 10 791.3569 | 11 474.4490 | 12 178.9934 | 12 905.0690 | 13 652.7333 | 14 422.0648 | 15 213.1416 | 16 026.0428 | 16 860.8470 |
| i | $1s(^2S)2p^2(^3P)\,^4P_{1/2}$ | $1s^22p\,^2P^o_{1/2}$ | 8 899.8970 | 9 523.4575 | 10 168.8080 | 10 836.0386 | 11 525.2486 | 12 236.5305 | 12 970.0054 | 13 725.7766 | 14 503.9698 | 15 304.7072 | 16 128.1223 | 16 974.3475 |
| j | $1s(^2S)2p^2(^1D)\,^2D_{5/2}$ | $1s^22p\,^2P^o_{3/2}$ | 8 931.6692 | 9 558.5808 | 10 207.8231 | 10 879.5393 | 11 573.8836 | 12 290.9977 | 13 031.0541 | 13 794.2059 | 14 580.6281 | 15 390.4972 | 16 223.9980 | 17 081.3154 |
| k | $1s(^2S)2p^2(^1D)\,^2D_{3/2}$ | $1s^22p\,^2P^o_{1/2}$ | 8 945.8910 | 9 573.7700 | 10 223.9435 | 10 896.5533 | 11 591.7467 | 12 309.6708 | 13 050.4943 | 13 814.3723 | 14 601.4811 | 15 411.9950 | 16 246.0988 | 17 103.9794 |
| l | $1s(^2S)2p^2(^1D)\,^2D_{3/2}$ | $1s^22p\,^2P^o_{3/2}$ | 8 916.2479 | 9 539.6280 | 10 184.8003 | 10 851.8718 | 11 540.9482 | 12 252.1343 | 12 985.5579 | 13 741.3298 | 14 519.5777 | 15 320.4285 | 16 144.0176 | 16 990.4791 |
| m | $1s(^2S)2p^2(^1S)\,^2S_{1/2}$ | $1s^22p\,^2P^o_{3/2}$ | 8 969.9459 | 9 598.1165 | 10 248.5896 | 10 921.5124 | 11 617.0381 | 12 335.3148 | 13 076.5170 | 13 840.8001 | 14 628.3459 | 15 439.3322 | 16 273.9462 | 17 132.3764 |
| n | $1s(^2S)2p^2(^1S)\,^2S_{1/2}$ | $1s^22p\,^2P^o_{1/2}$ | 8 999.5891 | 9 632.2586 | 10 287.7321 | 10 966.1943 | 11 667.8371 | 12 392.8517 | 13 141.4534 | 13 913.8433 | 14 710.2504 | 15 530.8979 | 16 376.0276 | 17 245.8772 |
| o | $1s2s^2\,^2S_{1/2}$ | $1s^22p\,^2P^o_{3/2}$ | 8 784.7267 | 9 399.6924 | 10 035.9103 | 10 693.4535 | 11 372.3778 | 12 072.7466 | 12 794.6404 | 13 538.1197 | 14 303.2659 | 15 090.1531 | 15 898.8630 | 16 729.4729 |
| p | $1s2s^2\,^2S_{1/2}$ | $1s^22p\,^2P^o_{1/2}$ | 8 814.3701 | 9 433.8345 | 10 075.0531 | 10 738.1346 | 11 423.1764 | 12 130.2836 | 12 859.5758 | 13 611.1636 | 14 385.1701 | 15 181.7194 | 16 000.9445 | 16 842.9742 |
| q | $1s(^2S)2s2p(^3P^o)\,^2P^o_{3/2}$ | $1s^22s\,^2S_{1/2}$ | 8 952.6613 | 9 580.2925 | 10 230.1937 | 10 902.5104 | 11 597.3969 | 12 315.0042 | 13 055.5006 | 13 819.0515 | 14 605.8312 | 15 416.0193 | 16 249.8058 | 17 107.3778 |
| r | $1s(^2S)2s2p(^3P^o)\,^2P^o_{1/2}$ | $1s^22s\,^2S_{1/2}$ | 8 933.5658 | 9 557.7975 | 10 203.8284 | 10 871.7586 | 11 561.6986 | 12 273.7529 | 13 008.0431 | 13 764.6845 | 14 543.8040 | 15 345.5267 | 16 169.9931 | 17 017.3322 |
| s | $1s(^2S)2s2p(^1P^o)\,^2P^o_{3/2}$ | $1s^22s\,^2S_{1/2}$ | 8 973.9115 | 9 602.7731 | 10 253.9608 | 10 927.6143 | 11 623.8887 | 12 342.9273 | 13 084.9036 | 13 849.9717 | 14 638.3113 | 15 450.0947 | 16 285.5128 | 17 144.7502 |
| t | $1s(^2S)2s2p(^1P^o)\,^2P^o_{1/2}$ | $1s^22s\,^2S_{1/2}$ | 8 969.0596 | 9 597.3899 | 10 248.0195 | 10 921.0911 | 11 616.7551 | 12 335.1590 | 13 076.4715 | 13 840.8511 | 14 628.4719 | 15 439.5091 | 16 274.1513 | 17 132.5847 |
| u | $1s(^2S)2s2p(^3P^o)\,^4P^o_{3/2}$ | $1s^22s\,^2S_{1/2}$ | 8 891.8872 | 9 514.7853 | 10 159.4806 | 10 826.0677 | 11 514.6519 | 12 225.3263 | 12 958.2134 | 13 713.4157 | 14 491.0592 | 15 291.2621 | 16 114.1577 | 16 959.8749 |
| v | $1s(^2S)2s2p(^3P^o)\,^4P^o_{1/2}$ | $1s^22s\,^2S_{1/2}$ | 8 885.9541 | 9 508.2524 | 10 152.3404 | 10 818.3208 | 11 506.3017 | 12 216.3835 | 12 948.6913 | 13 703.3350 | 14 480.4408 | 15 280.1339 | 16 102.5466 | 16 947.8127 |

| Key | Upper | Lower | 42 Mo | 43 Tc | 44 Ru | 45 Rh | 46 Pd | 47 Ag | 48 Cd | 49 In | 50 Sn | 51 Sb | 52 Te | 53 I |
|---|---|---|---|---|---|---|---|---|---|---|---|---|---|---|
| a | $1s(^2S)2p^2(^3P)\,^2P_{3/2}$ | $1s^22p\,^2P^o_{3/2}$ | 17 993.0333 | 18 899.6010 | 19 830.5993 | 20 786.2806 | 21 766.8505 | 22 772.5721 | 23 803.6696 | 24 860.4277 | 25 943.0938 | 27 051.9556 | 28 187.2789 | 29 349.4153 |
| b | $1s(^2S)2p^2(^3P)\,^2P_{3/2}$ | $1s^22p\,^2P^o_{1/2}$ | 18 118.9118 | 19 038.8776 | 19 984.3455 | 20 955.6342 | 21 953.0153 | 22 976.8141 | 24 027.3215 | 25 104.9019 | 26 209.8715 | 27 342.5937 | 28 503.4120 | 29 692.7743 |
| c | $1s(^2S)2p^2(^3P)\,^2P_{1/2}$ | $1s^22p\,^2P^o_{3/2}$ | 17 850.4869 | 18 742.5358 | 19 657.9085 | 20 596.7880 | 21 559.3198 | 22 545.6887 | 23 556.0529 | 24 590.6227 | 25 649.5646 | 26 733.0849 | 27 841.3654 | 28 974.6816 |
| d | $1s(^2S)2p^2(^3P)\,^2P_{1/2}$ | $1s^22p\,^2P^o_{1/2}$ | 17 976.3657 | 18 881.8133 | 19 811.6564 | 20 766.1437 | 21 745.4831 | 22 749.9285 | 23 779.7056 | 24 835.0957 | 25 916.3373 | 27 023.7184 | 28 157.4992 | 29 318.0459 |
| e | $1s(^2S)2p^2(^3P)\,^4P_{5/2}$ | $1s^22p\,^2P^o_{3/2}$ | 17 833.8155 | 18 725.7297 | 19 640.9273 | 20 579.5984 | 21 541.8865 | 22 527.9793 | 23 538.0350 | 24 572.2639 | 25 630.8416 | 26 713.9676 | 27 821.8293 | 28 954.6931 |
| f | $1s(^2S)2p^2(^3P)\,^4P_{3/2}$ | $1s^22p\,^2P^o_{3/2}$ | 17 820.9064 | 18 712.7064 | 19 627.8246 | 20 566.4487 | 21 528.7235 | 22 514.8392 | 23 524.9492 | 24 559.2681 | 25 617.9673 | 26 701.2474 | 27 809.2986 | 28 942.3780 |
| g | $1s(^2S)2p^2(^3P)\,^4P_{3/2}$ | $1s^22p\,^2P^o_{1/2}$ | 17 946.7877 | 18 851.9807 | 19 781.5706 | 20 735.8031 | 21 714.8889 | 22 719.0799 | 23 748.6002 | 24 803.7401 | 25 884.7390 | 26 991.8845 | 28 125.4342 | 29 285.7404 |
| h | $1s(^2S)2p^2(^3P)\,^4P_{1/2}$ | $1s^22p\,^2P^o_{3/2}$ | 17 717.6474 | 18 596.5549 | 19 497.6388 | 20 421.0263 | 21 366.8014 | 22 335.0805 | 23 325.9562 | 24 339.5674 | 25 376.0064 | 26 435.4066 | 27 517.8669 | 28 623.5772 |
| i | $1s(^2S)2p^2(^3P)\,^4P_{1/2}$ | $1s^22p\,^2P^o_{1/2}$ | 17 843.5275 | 18 735.8290 | 19 651.3873 | 20 590.3813 | 21 552.9635 | 22 539.3195 | 23 549.6055 | 24 584.0402 | 25 642.7796 | 26 726.0431 | 27 834.0025 | 28 966.9374 |
| j | $1s(^2S)2p^2(^1D)\,^2D_{5/2}$ | $1s^22p\,^2P^o_{3/2}$ | 17 962.6457 | 18 868.2189 | 19 798.2269 | 20 752.9143 | 21 732.4892 | 22 737.2162 | 23 767.3111 | 24 823.0677 | 25 904.7279 | 27 012.5730 | 28 146.8751 | 29 307.9830 |
| k | $1s(^2S)2p^2(^1D)\,^2D_{3/2}$ | $1s^22p\,^2P^o_{1/2}$ | 17 985.8370 | 18 891.9033 | 19 822.3623 | 20 777.4642 | 21 757.4196 | 22 762.4837 | 23 792.8846 | 24 848.9081 | 25 930.7987 | 27 038.8350 | 28 173.2902 | 29 334.5204 |
| l | $1s(^2S)2p^2(^1D)\,^2D_{3/2}$ | $1s^22p\,^2P^o_{3/2}$ | 17 859.9573 | 18 752.6277 | 19 668.6143 | 20 608.1095 | 21 571.2554 | 22 558.2452 | 23 569.2315 | 24 604.4330 | 25 664.0218 | 26 748.2012 | 27 857.1542 | 28 991.1588 |
| m | $1s(^2S)2p^2(^1S)\,^2S_{1/2}$ | $1s^22p\,^2P^o_{3/2}$ | 18 014.8241 | 18 921.5189 | 19 852.6551 | 20 808.4786 | 21 789.2020 | 22 795.0806 | 23 826.3496 | 24 883.2892 | 25 966.1418 | 27 075.2017 | 28 210.7269 | 29 373.0839 |
| n | $1s(^2S)2p^2(^1S)\,^2S_{1/2}$ | $1s^22p\,^2P^o_{1/2}$ | 18 140.7035 | 19 060.7947 | 20 006.3994 | 20 977.8335 | 21 975.3654 | 22 999.3230 | 24 050.0006 | 25 127.7572 | 26 232.9133 | 27 365.8348 | 28 526.8641 | 29 716.4446 |
| o | $1s2s^2\,^2S_{1/2}$ | $1s^22p\,^2P^o_{3/2}$ | 17 582.0742 | 18 456.7695 | 19 353.6340 | 20 272.7852 | 21 214.3002 | 22 178.3049 | 23 164.8873 | 24 174.1711 | 25 206.2599 | 26 261.2845 | 27 339.3319 | 28 440.5914 |
| p | $1s2s^2\,^2S_{1/2}$ | $1s^22p\,^2P^o_{1/2}$ | 17 707.9555 | 18 596.0444 | 19 507.3819 | 20 442.1371 | 21 400.4659 | 22 382.5463 | 23 388.5364 | 24 418.6437 | 25 473.0367 | 26 551.9160 | 27 655.4694 | 28 783.9525 |
| q | $1s(^2S)2s2p(^3P^o)\,^2P^o_{3/2}$ | $1s^22s\,^2S_{1/2}$ | 17 988.9320 | 18 894.7046 | 19 824.8790 | 20 779.7137 | 21 759.4090 | 22 764.2265 | 23 794.3915 | 24 850.1930 | 25 931.8671 | 27 039.7018 | 28 173.9624 | 29 335.0062 |
| r | $1s(^2S)2s2p(^3P^o)\,^2P^o_{1/2}$ | $1s^22s\,^2S_{1/2}$ | 17 887.6906 | 18 781.2361 | 19 698.1069 | 20 638.4826 | 21 602.5147 | 22 590.3910 | 23 602.2720 | 24 638.3657 | 25 698.8434 | 26 783.9171 | 27 893.7675 | 29 028.6679 |
| s | $1s(^2S)2s2p(^1P^o)\,^2P^o_{3/2}$ | $1s^22s\,^2S_{1/2}$ | 18 028.0104 | 18 935.5170 | 19 867.4637 | 20 824.1010 | 21 805.6382 | 22 812.3353 | 23 844.4173 | 24 902.1709 | 25 985.8427 | 27 095.7161 | 28 232.0603 | 29 395.2295 |
| t | $1s(^2S)2s2p(^1P^o)\,^2P^o_{1/2}$ | $1s^22s\,^2S_{1/2}$ | 18 015.0047 | 18 921.6430 | 19 852.6900 | 20 808.3948 | 21 788.9646 | 22 794.6573 | 23 825.6994 | 24 882.3703 | 25 964.9237 | 27 073.6349 | 28 208.7757 | 29 370.7042 |
| u | $1s(^2S)2s2p(^3P^o)\,^4P^o_{3/2}$ | $1s^22s\,^2S_{1/2}$ | 17 828.5532 | 18 720.3633 | 19 635.4279 | 20 573.9398 | 21 536.0343 | 22 521.8982 | 23 531.6957 | 24 565.6280 | 25 623.8653 | 26 706.6193 | 27 814.0587 | 28 946.4661 |
| v | $1s(^2S)2s2p(^3P^o)\,^4P^o_{1/2}$ | $1s^22s\,^2S_{1/2}$ | 17 816.0742 | 18 707.4998 | 19 622.2207 | 20 560.4188 | 21 522.2433 | 22 507.8785 | 23 517.4794 | 24 551.2632 | 25 609.3898 | 26 692.0730 | 27 799.4904 | 28 931.9165 |

References:
$Z = 6$–17: K$\alpha$ ($n = 2$): V.A. Yerokhin, A. Surzhykov & A. Müller (2017) PRA 96, 042505 – doi:10.1103/PhysRevA.96.042505
$Z = 18$–92: K$\alpha$ ($n = 2$): V.A. Yerokhin & A. Surzhykov (2018) JPCRD 47, 023105 – doi:10.1063/1.5034574

Compiled by N. Hell, LLNL. hell1@llnl.gov . Cite the quoted references when using these values!

Table 3: Energies in eV for transitions to the $n = 2$ shell of Li-like ions

| Key | Upper | Lower | 54 Xe | 55 Cs | 56 Ba | 57 La | 58 Ce | 59 Pr | 60 Nd | 61 Pm | 62 Sm | 63 Eu | 64 Gd | 65 Tb |
|---|---|---|---|---|---|---|---|---|---|---|---|---|---|---|
| a | $1s(^2S)2p^2(^3P)\,^2P_{3/2}$ | $1s^22p\,^2P^o_{3/2}$ | 30 538.6193 | 31 755.2624 | 32 999.6304 | 34 272.1028 | 35 573.0161 | 36 902.7609 | 38 261.6972 | 39 650.1764 | 41 068.4587 | 42 517.3764 | 43 997.1378 | 45 508.7419 |
| b | $1s(^2S)2p^2(^3P)\,^2P_{3/2}$ | $1s^22p\,^2P^o_{1/2}$ | 30 911.0134 | 32 158.5907 | 33 435.8917 | 34 743.3842 | 36 081.5134 | 37 450.7749 | 38 851.6247 | 40 284.5486 | 41 749.9072 | 43 248.6825 | 44 781.1722 | 46 348.5227 |
| c | $1s(^2S)2p^2(^3P)\,^2P_{1/2}$ | $1s^22p\,^2P^o_{3/2}$ | 30 133.1858 | 31 317.1601 | 32 526.7930 | 33 762.3693 | 35 024.1114 | 36 312.3076 | 37 627.1934 | 38 969.0243 | 40 337.9181 | 41 734.6029 | 43 159.1206 | 44 612.3532 |
| d | $1s(^2S)2p^2(^3P)\,^2P_{1/2}$ | $1s^22p\,^2P^o_{1/2}$ | 30 505.5809 | 31 720.4901 | 32 963.0539 | 34 233.6452 | 35 532.6037 | 36 860.3135 | 38 217.1282 | 39 603.3912 | 41 019.3680 | 42 465.8974 | 43 943.1524 | 45 452.1353 |
| e | $1s(^2S)2p^2(^3P)\,^4P_{5/2}$ | $1s^22p\,^2P^o_{3/2}$ | 30 112.7303 | 31 296.2037 | 32 505.3119 | 33 740.3368 | 35 001.5085 | 36 289.1166 | 37 603.3765 | 38 944.5677 | 40 312.8017 | 41 708.7839 | 43 132.6049 | 44 585.0806 |
| f | $1s(^2S)2p^2(^3P)\,^4P_{3/2}$ | $1s^22p\,^2P^o_{3/2}$ | 30 100.6676 | 31 284.4216 | 32 493.8538 | 33 729.2303 | 34 990.7808 | 36 278.7849 | 37 593.5025 | 38 935.1629 | 40 303.9036 | 41 700.4231 | 43 124.8036 | 44 577.8830 |
| g | $1s(^2S)2p^2(^3P)\,^4P_{3/2}$ | $1s^22p\,^2P^o_{1/2}$ | 30 473.0633 | 31 687.7537 | 32 930.1090 | 34 200.5090 | 35 499.2746 | 36 826.8000 | 38 183.4315 | 39 569.5303 | 40 985.3467 | 42 431.7150 | 43 908.8373 | 45 417.6866 |
| h | $1s(^2S)2p^2(^3P)\,^4P_{1/2}$ | $1s^22p\,^2P^o_{3/2}$ | 29 752.6135 | 30 905.1575 | 32 081.3040 | 33 281.2483 | 34 505.0972 | 35 753.0563 | 37 025.2286 | 38 321.7622 | 39 642.6712 | 40 988.5308 | 42 359.2889 | 43 755.6284 |
| i | $1s(^2S)2p^2(^3P)\,^4P_{1/2}$ | $1s^22p\,^2P^o_{1/2}$ | 30 125.0076 | 31 308.4848 | 32 517.5626 | 33 752.5255 | 35 013.5973 | 36 301.0698 | 37 615.1613 | 38 956.1312 | 40 324.1166 | 41 719.8292 | 43 143.3064 | 44 595.4242 |
| j | $1s(^2S)2p^2(^1D)\,^2D_{5/2}$ | $1s^22p\,^2P^o_{3/2}$ | 30 496.1566 | 31 711.7521 | 32 955.0721 | 34 226.4819 | 35 526.3319 | 36 854.9994 | 38 212.8407 | 39 600.2163 | 41 017.4003 | 42 465.1992 | 43 943.8221 | 45 454.2849 |
| k | $1s(^2S)2p^2(^1D)\,^2D_{3/2}$ | $1s^22p\,^2P^o_{1/2}$ | 30 522.7711 | 31 738.4110 | 32 981.7224 | 34 253.0998 | 35 552.8698 | 36 881.4208 | 38 239.1108 | 39 626.2887 | 41 043.2126 | 42 490.7257 | 43 969.0214 | 45 479.0947 |
| l | $1s(^2S)2p^2(^1D)\,^2D_{3/2}$ | $1s^22p\,^2P^o_{3/2}$ | 30 150.3768 | 31 335.0795 | 32 545.4660 | 33 781.8256 | 35 044.3759 | 36 333.4093 | 37 649.1768 | 38 991.9174 | 40 361.7651 | 41 759.4412 | 43 184.9920 | 44 639.3057 |
| m | $1s(^2S)2p^2(^1S)\,^2S_{1/2}$ | $1s^22p\,^2P^o_{3/2}$ | 30 562.5127 | 31 779.3877 | 33 023.9954 | 34 296.7234 | 35 597.8965 | 36 927.9198 | 38 287.1241 | 39 675.8830 | 41 094.4715 | 42 543.6955 | 44 023.7739 | 45 535.6848 |
| n | $1s(^2S)2p^2(^1S)\,^2S_{1/2}$ | $1s^22p\,^2P^o_{1/2}$ | 30 934.9068 | 32 182.7148 | 33 460.2552 | 34 768.0043 | 36 106.3953 | 37 475.9280 | 38 877.0496 | 40 310.2590 | 41 775.9179 | 43 274.9937 | 44 807.7947 | 46 375.4807 |
| o | $1s2s^2\,^2S_{1/2}$ | $1s^22p\,^2P^o_{3/2}$ | 29 565.1411 | 30 713.1584 | 31 884.7411 | 33 080.0691 | 34 299.2662 | 35 542.5046 | 36 809.9295 | 38 101.6677 | 39 417.7881 | 40 758.7479 | 42 124.5716 | 43 515.7936 |
| p | $1s2s^2\,^2S_{1/2}$ | $1s^22p\,^2P^o_{1/2}$ | 29 937.5333 | 31 116.4860 | 32 321.0005 | 33 551.3452 | 34 807.7604 | 36 090.5144 | 37 399.8592 | 38 736.0438 | 40 099.2383 | 41 490.0433 | 42 908.6046 | 44 355.5862 |
| q | $1s(^2S)2s2p(^3P^o)\,^2P^o_{3/2}$ | $1s^22s\,^2S_{1/2}$ | 30 523.0792 | 31 738.5410 | 32 981.6961 | 34 252.9105 | 35 552.5232 | 36 880.9161 | 38 238.4503 | 39 625.4655 | 41 042.2343 | 42 489.6045 | 43 967.7429 | 45 477.6434 |
| r | $1s(^2S)2s2p(^3P^o)\,^2P^o_{1/2}$ | $1s^22s\,^2S_{1/2}$ | 30 188.7790 | 31 374.3775 | 32 585.6595 | 33 822.9092 | 35 086.3555 | 36 376.2839 | 37 692.9344 | 39 036.5553 | 40 407.2917 | 41 805.8372 | 43 232.2749 | 44 687.4448 |
| s | $1s(^2S)2s2p(^1P^o)\,^2P^o_{3/2}$ | $1s^22s\,^2S_{1/2}$ | 30 585.4778 | 31 803.1663 | 33 048.5989 | 34 322.1490 | 35 624.1422 | 36 954.9856 | 38 315.0238 | 39 704.6235 | 41 124.0224 | 42 574.0964 | 44 055.0127 | 45 567.7671 |
| t | $1s(^2S)2s2p(^1P^o)\,^2P^o_{1/2}$ | $1s^22s\,^2S_{1/2}$ | 30 559.6577 | 31 776.0076 | 33 020.0464 | 34 292.1606 | 35 592.6642 | 36 921.9595 | 38 280.3860 | 39 668.3173 | 41 086.0011 | 42 534.2816 | 44 013.3500 | 45 524.1817 |
| u | $1s(^2S)2s2p(^3P^o)\,^4P^o_{3/2}$ | $1s^22s\,^2S_{1/2}$ | 30 103.9930 | 31 286.9241 | 32 495.4378 | 33 729.8176 | 34 990.2970 | 36 277.1608 | 37 590.6303 | 38 930.9695 | 40 298.2969 | 41 693.3415 | 43 116.1354 | 44 567.5475 |
| v | $1s(^2S)2s2p(^3P^o)\,^4P^o_{1/2}$ | $1s^22s\,^2S_{1/2}$ | 30 089.5055 | 31 272.5379 | 32 481.2039 | 33 715.7930 | 34 976.5172 | 36 263.6747 | 37 577.4940 | 38 918.2238 | 40 285.9885 | 41 681.5255 | 43 104.8779 | 44 556.8965 |

| Key | Upper | Lower | 66 Dy | 67 Ho | 68 Er | 69 Tm | 70 Yb | 71 Lu | 72 Hf | 73 Ta | 74 W | 75 Re | 76 Os | 77 Ir |
|---|---|---|---|---|---|---|---|---|---|---|---|---|---|---|
| a | $1s(^2S)2p^2(^3P)\,^2P_{3/2}$ | $1s^22p\,^2P^o_{3/2}$ | 47 051.2554 | 48 626.6509 | 50 234.6534 | 51 876.3856 | 53 551.4299 | 55 261.0049 | 57 006.2735 | 58 787.0861 | 60 604.1727 | 62 458.3823 | 64 349.9456 | 66 280.5148 |
| b | $1s(^2S)2p^2(^3P)\,^2P_{3/2}$ | $1s^22p\,^2P^o_{1/2}$ | 47 949.9735 | 49 587.6098 | 51 261.2912 | 52 972.3173 | 54 720.4613 | 56 507.0708 | 58 333.5365 | 60 199.8006 | 62 106.9038 | 64 055.7844 | 66 046.9880 | 68 082.3365 |
| c | $1s(^2S)2p^2(^3P)\,^2P_{1/2}$ | $1s^22p\,^2P^o_{3/2}$ | 46 093.2304 | 47 603.5647 | 49 142.9208 | 50 712.2496 | 52 311.0060 | 53 940.2259 | 55 600.8634 | 57 292.5357 | 59 015.8418 | 60 771.4142 | 62 559.2299 | 64 380.7539 |
| d | $1s(^2S)2p^2(^3P)\,^2P_{1/2}$ | $1s^22p\,^2P^o_{1/2}$ | 46 991.9423 | 48 564.5195 | 50 169.5658 | 51 808.2110 | 53 480.0535 | 55 186.3035 | 56 928.0895 | 58 705.2787 | 60 518.5630 | 62 368.8382 | 64 256.2984 | 66 182.5817 |
| e | $1s(^2S)2p^2(^3P)\,^4P_{5/2}$ | $1s^22p\,^2P^o_{3/2}$ | 46 065.1959 | 47 574.7588 | 49 113.2924 | 50 681.7973 | 52 279.6841 | 53 908.0248 | 55 567.7454 | 57 258.4831 | 58 980.8328 | 60 735.3928 | 62 522.2251 | 64 342.6988 |
| f | $1s(^2S)2p^2(^3P)\,^4P_{3/2}$ | $1s^22p\,^2P^o_{3/2}$ | 46 058.6246 | 47 568.8449 | 49 108.0790 | 50 677.3020 | 52 275.9368 | 53 905.0482 | 55 565.5788 | 57 257.1610 | 58 980.3558 | 60 735.8094 | 62 523.5494 | 64 344.9361 |
| g | $1s(^2S)2p^2(^3P)\,^4P_{3/2}$ | $1s^22p\,^2P^o_{1/2}$ | 46 957.3438 | 48 529.7899 | 50 134.7131 | 51 773.2504 | 53 444.9664 | 55 151.1260 | 56 892.7977 | 58 669.8875 | 60 483.0767 | 62 333.2177 | 64 220.5856 | 66 146.7783 |
| h | $1s(^2S)2p^2(^3P)\,^4P_{1/2}$ | $1s^22p\,^2P^o_{3/2}$ | 45 176.4194 | 46 623.2909 | 48 095.6537 | 49 594.2943 | 51 118.5245 | 52 669.1803 | 54 247.0333 | 55 851.5809 | 57 483.1497 | 59 142.1569 | 60 828.4509 | 62 543.1670 |
| i | $1s(^2S)2p^2(^3P)\,^4P_{1/2}$ | $1s^22p\,^2P^o_{1/2}$ | 46 075.1419 | 47 584.2352 | 49 122.2823 | 50 690.2514 | 52 287.5551 | 53 915.2450 | 55 574.2712 | 57 264.3012 | 58 985.8556 | 60 739.5882 | 62 525.5043 | 64 345.0029 |
| j | $1s(^2S)2p^2(^1D)\,^2D_{5/2}$ | $1s^22p\,^2P^o_{3/2}$ | 46 995.6294 | 48 569.8655 | 50 176.6721 | 51 817.1967 | 53 491.0363 | 55 199.3991 | 56 943.4371 | 58 722.9903 | 60 538.7751 | 62 391.6868 | 64 281.9509 | 66 211.2099 |
| k | $1s(^2S)2p^2(^1D)\,^2D_{3/2}$ | $1s^22p\,^2P^o_{1/2}$ | 47 020.0287 | 48 593.7940 | 50 200.0965 | 51 840.0541 | 53 513.2695 | 55 220.9603 | 56 964.2625 | 58 743.0505 | 60 558.0246 | 62 410.0280 | 64 299.3529 | 66 227.5851 |
| l | $1s(^2S)2p^2(^1D)\,^2D_{3/2}$ | $1s^22p\,^2P^o_{3/2}$ | 46 121.3162 | 47 632.8447 | 49 173.4626 | 50 744.1092 | 52 344.2437 | 53 974.8856 | 55 637.0416 | 57 330.3134 | 59 055.3082 | 60 812.6083 | 62 602.3153 | 64 425.7497 |
| m | $1s(^2S)2p^2(^1S)\,^2S_{1/2}$ | $1s^22p\,^2P^o_{3/2}$ | 47 078.5189 | 48 654.2629 | 50 262.5941 | 51 904.6618 | 53 580.0570 | 55 290.0101 | 57 035.6709 | 58 816.8425 | 60 634.2853 | 62 488.8858 | 64 380.8542 | 66 311.8167 |
| n | $1s(^2S)2p^2(^1S)\,^2S_{1/2}$ | $1s^22p\,^2P^o_{1/2}$ | 47 977.2305 | 49 615.2124 | 51 289.2401 | 53 000.6004 | 54 749.0950 | 56 536.0844 | 58 362.8904 | 60 229.5419 | 62 137.0027 | 64 086.3117 | 66 077.8936 | 68 113.6425 |
| o | $1s2s^2\,^2S_{1/2}$ | $1s^22p\,^2P^o_{3/2}$ | 44 931.6439 | 46 373.3645 | 47 840.5897 | 49 333.9460 | 50 852.9823 | 52 398.3979 | 53 970.7739 | 55 569.8375 | 57 195.9078 | 58 849.3385 | 60 530.1449 | 62 239.1575 |
| p | $1s2s^2\,^2S_{1/2}$ | $1s^22p\,^2P^o_{1/2}$ | 45 830.3648 | 47 334.3177 | 48 867.2385 | 50 429.8876 | 52 022.0259 | 53 644.4582 | 55 298.0245 | 56 982.5627 | 58 698.6361 | 60 446.7773 | 62 227.1935 | 64 040.9617 |
| q | $1s(^2S)2s2p(^3P^o)\,^2P^o_{3/2}$ | $1s^22s\,^2S_{1/2}$ | 47 018.3882 | 48 591.9657 | 50 198.0844 | 51 837.8216 | 53 510.8213 | 55 218.2796 | 56 961.3314 | 58 739.8221 | 60 554.5050 | 62 406.2269 | 64 295.1849 | 66 223.0572 |
| r | $1s(^2S)2s2p(^3P^o)\,^2P^o_{1/2}$ | $1s^22s\,^2S_{1/2}$ | 46 170.3166 | 47 682.6906 | 49 224.1243 | 50 795.6050 | 52 396.5378 | 54 028.0181 | 55 690.9220 | 57 384.9752 | 59 110.7177 | 60 868.7661 | 62 659.1687 | 64 483.3489 |
| s | $1s(^2S)2s2p(^1P^o)\,^2P^o_{3/2}$ | $1s^22s\,^2S_{1/2}$ | 47 111.4702 | 48 688.0619 | 50 297.2575 | 51 940.1918 | 53 616.4810 | 55 327.2908 | 57 073.8461 | 58 855.9035 | 60 674.2841 | 62 529.7929 | 64 422.6699 | 66 354.5811 |
| t | $1s(^2S)2s2p(^1P^o)\,^2P^o_{1/2}$ | $1s^22s\,^2S_{1/2}$ | 47 065.8837 | 48 640.4052 | 50 247.4591 | 51 888.1527 | 53 562.1181 | 55 270.5631 | 57 014.6097 | 58 794.1111 | 60 609.8018 | 62 462.5043 | 64 352.5174 | 66 281.4006 |
| u | $1s(^2S)2s2p(^3P^o)\,^4P^o_{3/2}$ | $1s^22s\,^2S_{1/2}$ | 46 046.5138 | 47 554.8690 | 49 092.1539 | 50 659.3287 | 52 255.8208 | 53 882.6757 | 55 540.8366 | 57 229.9652 | 58 950.6018 | 60 703.3964 | 62 488.3504 | 64 306.8567 |
| v | $1s(^2S)2s2p(^3P^o)\,^4P^o_{1/2}$ | $1s^22s\,^2S_{1/2}$ | 46 036.5288 | 47 545.6049 | 49 083.6609 | 50 651.6712 | 52 249.0822 | 53 876.9157 | 55 536.1097 | 57 226.3463 | 58 948.1634 | 60 702.1779 | 62 488.4764 | 64 308.3576 |

References:
$Z = 6$–17: K$\alpha$ ($n = 2$): V.A. Yerokhin, A. Surzhykov & A. Müller (2017) PRA 96, 042505 – doi:10.1103/PhysRevA.96.042505
$Z = 18$–92: K$\alpha$ ($n = 2$): V.A. Yerokhin & A. Surzhykov (2018) JPCRD 47, 023105 – doi:10.1063/1.5034574

Compiled by N. Hell, LLNL. hell1@llnl.gov . Cite the quoted references when using these values!

Table 3: Energies in eV for transitions to the $n = 2$ shell of Li-like ions

| Key | Upper | Lower | 78 Pt | 79 Au | 80 Hg | 81 Tl | 82 Pb | 83 Bi | 84 Po | 85 At | 86 Rn | 87 Fr | 88 Ra | 89 Ac |
|---|---|---|---|---|---|---|---|---|---|---|---|---|---|---|
| a | $1s(^2S)2p^2(^3P)\,^2P_{3/2}$ | $1s^22p\,^2P^o_{3/2}$ | 68 249.8231 | 70 259.4994 | 72 309.5936 | 74 401.7537 | 76 536.1015 | 78 714.1287 | 80 936.8712 | 83 204.9521 | 85 515.7210 | 87 877.8954 | 90 288.1275 | 92 751.1002 |
| b | $1s(^2S)2p^2(^3P)\,^2P_{3/2}$ | $1s^22p\,^2P^o_{1/2}$ | 70 161.8106 | 72 287.2492 | 74 458.9915 | 76 678.9565 | 78 947.4948 | 81 266.4755 | 83 637.1402 | 86 060.5141 | 88 534.2138 | 91 067.3561 | 93 657.0239 | 96 308.2057 |
| c | $1s(^2S)2p^2(^3P)\,^2P_{1/2}$ | $1s^22p\,^2P^o_{3/2}$ | 66 235.4749 | 68 124.7206 | 70 048.3216 | 72 007.6003 | 74 002.4355 | 76 034.0715 | 78 103.1360 | 80 209.8910 | 82 351.5414 | 84 536.1601 | 86 760.1779 | 89 027.7920 |
| d | $1s(^2S)2p^2(^3P)\,^2P_{1/2}$ | $1s^22p\,^2P^o_{1/2}$ | 68 147.4494 | 70 152.4814 | 72 197.7156 | 74 284.8266 | 76 413.8351 | 78 586.4043 | 80 803.3650 | 83 065.4789 | 85 370.0460 | 87 725.6204 | 90 129.0306 | 92 584.9414 |
| e | $1s(^2S)2p^2(^3P)\,^4P_{5/2}$ | $1s^22p\,^2P^o_{3/2}$ | 66 196.3272 | 68 084.4676 | 70 006.9499 | 71 965.0939 | 73 958.7773 | 75 989.1949 | 78 056.9639 | 80 162.4909 | 82 302.8336 | 84 486.1012 | 86 708.7853 | 88 974.9557 |
| f | $1s(^2S)2p^2(^3P)\,^4P_{3/2}$ | $1s^22p\,^2P^o_{3/2}$ | 66 199.5436 | 68 088.6927 | 70 012.1681 | 71 971.3601 | 73 966.0574 | 75 997.5790 | 78 066.4496 | 80 173.1174 | 82 314.5816 | 84 499.0567 | 86 722.8561 | 88 990.2826 |
| g | $1s(^2S)2p^2(^3P)\,^4P_{3/2}$ | $1s^22p\,^2P^o_{1/2}$ | 68 111.5097 | 70 116.4185 | 72 161.5778 | 74 248.5262 | 76 377.4478 | 78 549.8598 | 80 766.7292 | 83 028.6542 | 85 333.0294 | 87 688.5178 | 90 091.7661 | 92 547.4149 |
| h | $1s(^2S)2p^2(^3P)\,^4P_{1/2}$ | $1s^22p\,^2P^o_{3/2}$ | 64 285.6505 | 66 056.9464 | 67 856.6435 | 69 685.7941 | 71 543.9705 | 73 432.1270 | 75 350.5911 | 77 299.3396 | 79 275.3048 | 81 285.9226 | 83 327.4739 | 85 403.6238 |
| i | $1s(^2S)2p^2(^3P)\,^4P_{1/2}$ | $1s^22p\,^2P^o_{1/2}$ | 66 197.6349 | 68 084.6919 | 70 006.0407 | 71 962.9636 | 73 955.3804 | 75 984.4447 | 78 050.8707 | 80 154.8727 | 82 293.7654 | 84 475.3944 | 86 696.3558 | 88 960.7191 |
| j | $1s(^2S)2p^2(^1D)\,^2D_{5/2}$ | $1s^22p\,^2P^o_{3/2}$ | 68 179.1903 | 70 187.5085 | 72 236.2464 | 74 326.9992 | 76 459.9221 | 78 636.5466 | 80 857.8536 | 83 124.4554 | 85 433.7545 | 87 794.4487 | 90 203.1272 | 92 664.5872 |
| k | $1s(^2S)2p^2(^1D)\,^2D_{3/2}$ | $1s^22p\,^2P^o_{1/2}$ | 68 194.4904 | 70 201.6962 | 72 249.2114 | 74 338.6752 | 76 470.2498 | 78 645.4253 | 80 865.1841 | 83 130.1961 | 85 437.8167 | 87 796.6246 | 90 203.4554 | 92 662.8558 |
| l | $1s(^2S)2p^2(^1D)\,^2D_{3/2}$ | $1s^22p\,^2P^o_{3/2}$ | 66 282.5345 | 68 173.9418 | 70 099.8079 | 72 061.5056 | 74 058.8391 | 76 093.1022 | 78 164.9316 | 80 274.6510 | 82 419.3141 | 84 607.1737 | 86 834.6139 | 89 105.7233 |
| m | $1s(^2S)2p^2(^1S)\,^2S_{1/2}$ | $1s^22p\,^2P^o_{3/2}$ | 68 281.5466 | 70 291.6445 | 72 342.1651 | 74 434.7184 | 76 569.4718 | 78 747.9753 | 80 971.1761 | 83 239.6422 | 85 550.9482 | 87 913.5376 | 90 324.2386 | 92 787.6810 |
| n | $1s(^2S)2p^2(^1S)\,^2S_{1/2}$ | $1s^22p\,^2P^o_{1/2}$ | 70 193.5088 | 72 319.3788 | 74 491.5594 | 76 711.9294 | 78 980.8883 | 81 300.3140 | 83 671.4010 | 86 095.2351 | 88 569.3782 | 91 103.0223 | 93 693.1193 | 96 344.8017 |
| o | $1s2s^2\,^2S_{1/2}$ | $1s^22p\,^2P^o_{3/2}$ | 63 975.9948 | 65 741.5711 | 67 535.6260 | 69 359.0344 | 71 211.5364 | 73 094.0211 | 75 006.6991 | 76 949.7922 | 78 921.2122 | 80 926.4112 | 82 962.7626 | 85 033.0906 |
| p | $1s2s^2\,^2S_{1/2}$ | $1s^22p\,^2P^o_{1/2}$ | 65 887.9902 | 67 769.3332 | 69 685.0108 | 71 636.1934 | 73 622.9410 | 75 646.3390 | 77 706.9750 | 79 805.3264 | 81 939.7061 | 84 115.8222 | 86 331.6144 | 88 590.1991 |
| q | $1s(^2S)2s2p(^3P^o)\,^2P^o_{3/2}$ | $1s^22s\,^2S_{1/2}$ | 68 189.5771 | 70 196.3304 | 72 243.4018 | 74 332.3465 | 76 463.3644 | 78 637.9930 | 80 857.1681 | 83 121.5018 | 85 428.4565 | 87 786.5541 | 90 192.5627 | 92 651.1534 |
| r | $1s(^2S)2s2p(^3P^o)\,^2P^o_{1/2}$ | $1s^22s\,^2S_{1/2}$ | 66 340.7698 | 68 232.8083 | 70 159.3093 | 72 121.6163 | 74 119.4937 | 76 154.3295 | 78 226.6277 | 80 336.8086 | 82 481.9853 | 84 670.2108 | 86 897.9698 | 89 169.4878 |
| s | $1s(^2S)2s2p(^1P^o)\,^2P^o_{3/2}$ | $1s^22s\,^2S_{1/2}$ | 68 325.2711 | 70 336.3461 | 72 387.8230 | 74 481.4012 | 76 617.1198 | 78 796.6713 | 81 020.8612 | 83 290.4166 | 85 602.7501 | 87 966.4311 | 90 378.1630 | 92 842.7802 |
| t | $1s(^2S)2s2p(^1P^o)\,^2P^o_{1/2}$ | $1s^22s\,^2S_{1/2}$ | 68 248.9590 | 70 256.7921 | 72 304.9128 | 74 394.9679 | 76 527.0785 | 78 702.8363 | 80 923.1363 | 83 188.5947 | 85 496.6738 | 87 855.9761 | 90 263.1494 | 92 722.9380 |
| u | $1s(^2S)2s2p(^3P^o)\,^4P^o_{3/2}$ | $1s^22s\,^2S_{1/2}$ | 66 158.4261 | 68 044.4488 | 69 964.6794 | 71 920.5098 | 73 911.6896 | 75 939.5802 | 78 004.7113 | 80 107.4341 | 82 244.9077 | 84 425.1199 | 86 644.6149 | 88 907.5161 |
| v | $1s(^2S)2s2p(^3P^o)\,^4P^o_{1/2}$ | $1s^22s\,^2S_{1/2}$ | 66 161.4269 | 68 049.0424 | 69 970.9180 | 71 928.4791 | 73 921.5166 | 75 951.3031 | 78 018.5043 | 80 123.3788 | 82 263.0793 | 84 445.6481 | 86 667.6302 | 88 933.1528 |

| Key | Upper | Lower | 90 Th | 91 Pa | 92 U | 93 Np | 94 Pu | 95 Am | 96 Cm | 97 Bk | 98 Cf | 99 Es | 100 Fm | 101 Md |
|---|---|---|---|---|---|---|---|---|---|---|---|---|---|---|
| a | $1s(^2S)2p^2(^3P)\,^2P_{3/2}$ | $1s^22p\,^2P^o_{3/2}$ | 95 259.1752 | 97 827.5557 | 100 439.6402 | | | | | | | | | |
| b | $1s(^2S)2p^2(^3P)\,^2P_{3/2}$ | $1s^22p\,^2P^o_{1/2}$ | 99 013.6492 | 101 789.2549 | 104 618.3924 | | | | | | | | | |
| c | $1s(^2S)2p^2(^3P)\,^2P_{1/2}$ | $1s^22p\,^2P^o_{3/2}$ | 91 330.9936 | 93 684.4113 | 96 071.2679 | | | | | | | | | |
| d | $1s(^2S)2p^2(^3P)\,^2P_{1/2}$ | $1s^22p\,^2P^o_{1/2}$ | 95 085.5215 | 97 646.1125 | 100 250.0899 | | | | | | | | | |
| e | $1s(^2S)2p^2(^3P)\,^4P_{5/2}$ | $1s^22p\,^2P^o_{3/2}$ | 91 276.7329 | 93 628.7332 | 96 014.0558 | | | | | | | | | |
| f | $1s(^2S)2p^2(^3P)\,^4P_{3/2}$ | $1s^22p\,^2P^o_{3/2}$ | 91 293.3337 | 93 646.4836 | 96 033.1685 | | | | | | | | | |
| g | $1s(^2S)2p^2(^3P)\,^4P_{3/2}$ | $1s^22p\,^2P^o_{1/2}$ | 95 047.8354 | 97 608.2140 | 100 211.9254 | | | | | | | | | |
| h | $1s(^2S)2p^2(^3P)\,^4P_{1/2}$ | $1s^22p\,^2P^o_{3/2}$ | 87 506.2539 | 89 648.9279 | 91 815.3228 | | | | | | | | | |
| i | $1s(^2S)2p^2(^3P)\,^4P_{1/2}$ | $1s^22p\,^2P^o_{1/2}$ | 91 260.7426 | 93 610.6361 | 95 994.1331 | | | | | | | | | |
| j | $1s(^2S)2p^2(^1D)\,^2D_{5/2}$ | $1s^22p\,^2P^o_{3/2}$ | 95 171.0637 | 97 737.8671 | 100 348.3490 | | | | | | | | | |
| k | $1s(^2S)2p^2(^1D)\,^2D_{3/2}$ | $1s^22p\,^2P^o_{1/2}$ | 95 167.1920 | 97 731.7807 | 100 339.9030 | | | | | | | | | |
| l | $1s(^2S)2p^2(^1D)\,^2D_{3/2}$ | $1s^22p\,^2P^o_{3/2}$ | 91 412.7416 | 93 770.0740 | 96 161.0549 | | | | | | | | | |
| m | $1s(^2S)2p^2(^1S)\,^2S_{1/2}$ | $1s^22p\,^2P^o_{3/2}$ | 95 296.2233 | 97 865.0840 | 100 477.6525 | | | | | | | | | |
| n | $1s(^2S)2p^2(^1S)\,^2S_{1/2}$ | $1s^22p\,^2P^o_{1/2}$ | 99 050.7479 | 101 826.7906 | 104 656.4538 | | | | | | | | | |
| o | $1s2s^2\,^2S_{1/2}$ | $1s^22p\,^2P^o_{3/2}$ | 87 131.8638 | 89 268.6800 | 91 432.5609 | | | | | | | | | |
| p | $1s2s^2\,^2S_{1/2}$ | $1s^22p\,^2P^o_{1/2}$ | 90 886.3786 | 93 230.3846 | 95 611.3416 | | | | | | | | | |
| q | $1s(^2S)2s2p(^3P^o)\,^2P^o_{3/2}$ | $1s^22s\,^2S_{1/2}$ | 95 154.6294 | 97 718.2239 | 100 325.4506 | | | | | | | | | |
| r | $1s(^2S)2s2p(^3P^o)\,^2P^o_{1/2}$ | $1s^22s\,^2S_{1/2}$ | 91 476.7471 | 93 834.3706 | 96 225.6861 | | | | | | | | | |
| s | $1s(^2S)2s2p(^1P^o)\,^2P^o_{3/2}$ | $1s^22s\,^2S_{1/2}$ | 95 352.4363 | 97 922.4359 | 100 536.2333 | | | | | | | | | |
| t | $1s(^2S)2s2p(^1P^o)\,^2P^o_{1/2}$ | $1s^22s\,^2S_{1/2}$ | 95 227.6410 | 97 792.4473 | 100 400.9248 | | | | | | | | | |
| u | $1s(^2S)2s2p(^3P^o)\,^4P^o_{3/2}$ | $1s^22s\,^2S_{1/2}$ | 91 205.8944 | 93 554.2691 | 95 935.9735 | | | | | | | | | |
| v | $1s(^2S)2s2p(^3P^o)\,^4P^o_{1/2}$ | $1s^22s\,^2S_{1/2}$ | 91 234.2166 | 93 585.4815 | 95 970.1328 | | | | | | | | | |

References:
$Z$ = 6–17: K$\alpha$ ($n = 2$): V.A. Yerokhin, A. Surzhykov & A. Müller (2017) PRA 96, 042505 – doi:10.1103/PhysRevA.96.042505
$Z$ = 18–92: K$\alpha$ ($n = 2$): V.A. Yerokhin & A. Surzhykov (2018) JPCRD 47, 023105 – doi:10.1063/1.5034574

Compiled by N. Hell, LLNL. hell1@llnl.gov . Cite the quoted references when using these values!

Table 4: E models (measurement, theory, or combination) for lineshape and position of K-shell lines in neutrals (here: solid targets). Listed are energy (eV), Lorentzian width Γ (FWHM in eV), and relative amplitude (area $\cdot 2/\pi/\Gamma$) of the model components. Also listed are the normalized intensities (area) of the components with $\sum_i I_i = 1$. Note that the exact line shapes depend on source conditions (solid, gas; molecules causing chemical shifts), excitation process, and excitation energy.

| $Z$ | | Trans. Type | Energy [eV] | FWHM Γ [eV] | Amplitude area $\cdot 2/\pi/\Gamma$ | Intensity (normalized) | References |
|---|---|---|---|---|---|---|---|
| 9 | F | K $\alpha$ | 676.8<br>676.8 | 0.20<br>0.20 | 1.000000<br>0.500000 | 0.66667<br>0.33333 | [theory] Positions from Bearden (1967) in Zschornack (2007). Faked 2 Lorentzian mode. Only F Ka is in Bearden; width extrapolated from Krause & Oliver (1979) |
| 11 | Na | K $\alpha$ | 1 040.98<br>1 040.98 | 0.30<br>0.30 | 1.000000<br>0.500000 | 0.66667<br>0.33333 | [theory] Positions from Bearden (1967) in Zschornack (2007). 2 Lorentzian model (widths) from Krause & Oliver (1979). |
| 12 | Mg | K $\alpha$ | 1 253.687<br>1 253.436 | 0.36<br>0.36 | 1.000000<br>0.500000 | 0.66667<br>0.33333 | [mixed] Positions from Schweppe et al. 1994. 2 Lorentzian model (widths) from Krause & Oliver (1979). |
| 13 | Al | K $\alpha$ | 1 486.708<br>1 486.295 | 0.43<br>0.43 | 1.000000<br>0.503300 | 0.66520<br>0.33480 | [theory] Positions from Bearden (1967) in Zschornack (2007). 2 Lorentzian model (widths) from Krause & Oliver (1979). Relative intensities from Scofield (1974a). |
| | | K $\alpha$ | 1 486.9<br>1 486.5<br>1 496.4<br>1 498.4<br>1 492.3 | 0.43<br>0.43<br>0.960<br>1.252<br>1.340 | 1.000000<br>0.503300<br>0.053750<br>0.020607<br>0.006418 | 0.61036<br>0.30719<br>0.04947<br>0.02474<br>0.00825 | [theory] K$\alpha_1$, K$\alpha_2$ widths from Krause & Oliver (1979) and intensities from Scofield (1974a). Positions from Fischer & Baun (1965). K$\alpha_3$, K$\alpha_4$ (rows 3&4) widths from Nordfors (1955). Ka$\alpha'$ (row 5) width from Wollman et al. (2000). Because the satellite lines change based on details of the x-ray generator, their values may be different from the values listed in the table. If fitting for gain or line-spread function parameters, especially with a resolution $\lesssim 4$ eV, we suggest using the 2-Lorentzian model, which has the better energies. |
| 14 | Si | K $\alpha$ | 1 739.98<br>1 739.38 | 0.49<br>0.49 | 1.000000<br>0.503700 | 0.66503<br>0.33497 | [theory] Positions from Bearden (1967) in Zschornack (2007). 2 Lorentzian model (widths) from Krause & Oliver (1979). Relative intensities from Scofield (1974a). |
| 15 | P | K $\alpha$ | 2 013.7<br>2 012.7 | 0.57<br>0.56 | 1.000000<br>0.513814 | 0.66454<br>0.33546 | [theory] Positions from Bearden (1967). 2 Lorentzian model (widths) from Krause & Oliver (1979). Relative intensities from Scofield (1974a). |
| 16 | S | K $\alpha$ | 2 307.84<br>2 306.64 | 0.65<br>0.64 | 1.000000<br>0.513195 | 0.66432<br>0.33568 | [theory] Positions from Bearden (1967). 2 Lorentzian model (widths) from Krause & Oliver (1979). Relative intensities from Scofield (1974a). |
| 17 | Cl | K $\alpha$ | 2 622.39<br>2 620.78 | 0.72<br>0.72 | 1.000000<br>0.505600 | 0.66419<br>0.33581 | [theory] Positions from Bearden (1967) in Zschornack (2007). 2 Lorentzian model (widths) from Krause & Oliver (1979). Relative intensities from Scofield (1974a). |
| 18 | Ar | K $\alpha$ | 2 957.70<br>2 955.63 | 0.81<br>0.80 | 1.000000<br>0.511211 | 0.66450<br>0.33550 | [theory] Positions from Bearden (1967) in Zschornack (2007). 2 Lorentzian model (widths) from Krause & Oliver (1979). Relative intensities from Scofield (1974a). |
| 19 | K | K $\alpha$ | 3 313.80<br>3 311.10 | 0.89<br>0.89 | 1.000000<br>0.505500 | 0.66423<br>0.33577 | [theory] Positions from Bearden (1967) in Zschornack (2007). 2 Lorentzian model (widths) from Krause & Oliver (1979). Relative intensities from Scofield (1974b). |
| 20 | Ca | K $\alpha$ | 3 691.687<br>3 692.682<br>3 688.101<br>3 688.849<br>3 694.536 | 1.023<br>2.09<br>0.957<br>1.743<br>2.43 | 1.000<br>0.070<br>0.501<br>0.059<br>0.030 | 0.56076<br>0.08019<br>0.26260<br>0.05664<br>0.03981 | [empirical] Ito et al. (2016). Using FWHM column (uncorrected widths) for the satellite lines (2,4,5) and the CF column (corrected widths) for lines 1 and 3. Measurement done with $CaF_2$. |
| 21 | Sc | K $\alpha$ | 4 090.592<br>4 089.38<br>4 085.765<br>4 086.29<br>4 093.484 | 1.243<br>2.291<br>1.358<br>3.660<br>1.742 | 1.000<br>0.087<br>0.387<br>0.086<br>0.042 | 0.52790<br>0.08499<br>0.22330<br>0.13303<br>0.03078 | [empirical] Ito et al. (2016). [empirical] Using FWHM column (uncorrected widths) for the satellite lines (2,4,5) and the CF column (corrected widths) for lines 1 and 3. |
| 22 | Ti | K $\alpha$ | 4 510.918<br>4 509.954<br>4 507.763<br>4 514.002<br>4 504.910<br>4 503.088 | 1.37<br>2.22<br>3.75<br>1.70<br>1.88<br>4.49 | 1.0000<br>0.1337<br>0.0480<br>0.0301<br>0.4417<br>0.0104 | 0.49368<br>0.10696<br>0.06486<br>0.01844<br>0.29923<br>0.01683 | [empirical] 6 Lorentzian model from Chantler et al. (2006) on data from Kawai et al. (1994). The Lorentzian amplitudes were computed by us from integrated intensity reported in the paper. The Gaussian width from the Voigt fit was 0.11 eV. |
| | | K $\alpha$ ($TiO_2$) | 4 510.25<br>4 504.5 | 1.16<br>1.18 | 1.0<br>0.4 | 0.71078<br>0.28922 | [mixed] Positions from Bearden (1967). 2 Lorentzian model (widths) from Krause & Oliver (1979). Chemshift and K$\alpha_1$/K$\alpha_2$ ratio from Kavcic et al. 2005. |
| 23 | V | K $\alpha$ | 4 952.237<br>4 950.656<br>4 948.266<br>4 955.269<br>4 944.672<br>4 943.014 | 1.45<br>2.00<br>1.81<br>1.76<br>2.94<br>3.09 | 1.0000<br>0.1773<br>0.0532<br>0.0322<br>0.3592<br>0.0164 | 0.47319<br>0.11572<br>0.03142<br>0.01849<br>0.34463<br>0.01654 | [empirical] 6 Lorentzian model from Chantler et al. (2006). The Lorentzian amplitudes were computed by us from integrated intensity reported in the paper. The Gaussian width from the Voigt fit was 1.99 eV. |
| 24 | Cr | K $\alpha$ | 5 414.874<br>5 414.099<br>5 412.745<br>5 410.583<br>5 418.304<br>5 405.551<br>5 403.986 | 1.457<br>1.760<br>3.138<br>5.149<br>1.988<br>2.224<br>4.740 | 0.822<br>0.237<br>0.085<br>0.045<br>0.015<br>0.386<br>0.036 | 0.37800<br>0.13200<br>0.08400<br>0.07300<br>0.00900<br>0.27100<br>0.05400 | [empirical] 7 Lorentzian model from Hölzer et al. (1997). |
| | | K $\beta$ | 5 947.00<br>5 935.31<br>5 946.24<br>5 942.04<br>5 944.93 | 1.70<br>15.98<br>1.90<br>6.69<br>3.37 | 0.670<br>0.055<br>0.337<br>0.082<br>0.151 | 0.30700<br>0.23600<br>0.17200<br>0.14800<br>0.13700 | [empirical] 5 Lorentzian model from Hölzer et al. (1997) |

References:

- Bearden 1967, Rev. Modern Physics 39, 78 — doi:10.1103/RevModPhys.39.78 — https://ui.adsabs.harvard.edu/abs/1967RvMP...39...78B
- Chantler et al. 2006, PRA 73, 012508 — doi:10.1103/PhysRevA.73.012508 — https://ui.adsabs.harvard.edu/abs/2006PhRvA..73a2508C
- Fischer & Baun 1965, J. Applied Physics 36, 534 — doi:10.1063/1.1714025 — https://ui.adsabs.harvard.edu/abs/1965JAP....36..534F
- Hölzer et al. 1997, PRA 56, 4554 — doi:10.1103/PhysRevA.56.4554 — http://adsabs.harvard.edu/abs/1997PhRvA..56.4554H
- Ito et al. 2015, JQSRT 151, 295 — doi:10.1016/j.jqsrt.2014.10.013 — https://ui.adsabs.harvard.edu/abs/2015JQSRT.151..295I
- Ito et al. 2016, PRA 94, 042506 — doi:10.1103/PhysRevA.94.042506 — https://ui.adsabs.harvard.edu/abs/2016PhRvA..94d2506I
- Kavcic et al. 2005, X-ray Spectrometry 34, 310 — doi:10.1002/xrs.822 — https://ui.adsabs.harvard.edu/abs/2005XRS....34..310K
- Kawai et al. 1994, Spectrochimica Acta B 49, 725 — doi:10.1016/0584-8547(94)80064-2 — https://ui.adsabs.harvard.edu/abs/1994AcSpB..49..725K
- Krause & Oliver 1979, JPCRD 8, 329 — doi:10.1063/1.555595 — http://adsabs.harvard.edu/abs/1979JPCRD...8..329K
- Nordfors 1955, Phys. Soc. A 68, 654 — doi:10.1088/0370-1298/68/7/416 — https://ui.adsabs.harvard.edu/abs/1955PPSA...68..654N
- Schweppe et al. 1994, JESRP 67, 463 doi:10.1016/0368-2048(93)02059-U — https://ui.adsabs.harvard.edu/abs/1994JESRP..67..463S
- Scofield 1974a, PRA 9, 1041 — doi:10.1103/PhysRevA.9.1041 — https://ui.adsabs.harvard.edu/abs/1974PhRvA...9.1041S
- Scofield 1974b, ADNDT 14, 121 — doi:10.1016/S0092-640X(74)80019-7 — https://ui.adsabs.harvard.edu/abs/1974ADNDT..14..121S
- Wollman et al. 2000, NIMPRA 444, 145 — 10.1016/S0168-9002(99)01351-0 — https://ui.adsabs.harvard.edu/abs/2000NIMPA.444..145W
- Zschornack 2007, Handbook of X-ray Data,Springer — doi:10.1007/978-3-540-28619-6 — https://link.springer.com/book/10.1007/978-3-540-28619-6

Compiled by the NASA/GSFC calorimeter group (contact: F.S. Porter, frederick.s.porter@nasa.gov ; M.C. Witthoeft, michael.c.witthoeft@nasa.gov ).

Cite the quoted references when using these values!

Table 4: E models (measurement, theory, or combination) for lineshape and position of K-shell lines in neutrals (here: solid targets). Listed are energy (eV), Lorentzian width Γ (FWHM in eV), and relative amplitude (area $\cdot 2/\pi/\Gamma$) of the model components. Also listed are the normalized intensities (area) of the components with $\sum_i I_i = 1$. Note that the exact line shapes depend on source conditions (solid, gas; molecules causing chemical shifts), excitation process, and excitation energy.

| $Z$ | | Trans. Type | Energy [eV] | FWHM Γ [eV] | Amplitude area $\cdot 2/\pi/\Gamma$ | Intensity (normalized) | References |
|---|---|---|---|---|---|---|---|
| 25 | Mn | K α | 5 898.882 | 1.71450 | 0.784 | 0.35200 | [empirical] 8 Lorentzian model from Hölzer et al. (1997). Deconvolved spectra re-fit by F. S. Porter 11/30/2004, correcting several errors in the paper. |
| | | | 5 897.898 | 2.04420 | 0.263 | 0.14100 | |
| | | | 5 894.864 | 4.49850 | 0.067 | 0.07900 | |
| | | | 5 896.566 | 2.66160 | 0.095 | 0.06600 | |
| | | | 5 899.444 | 0.97669 | 0.071 | 0.01800 | |
| | | | 5 902.712 | 1.55280 | 0.011 | 0.00400 | |
| | | | 5 887.772 | 2.36040 | 0.369 | 0.22800 | |
| | | | 5 886.528 | 4.21680 | 0.100 | 0.11100 | |
| | | K β | 6 490.89 | 1.83 | 0.608 | 0.25400 | [empirical] 5 Lorentzian model from Hölzer et al. (1997). |
| | | | 6 486.31 | 9.40 | 0.109 | 0.23400 | |
| | | | 6 477.73 | 13.22 | 0.077 | 0.23400 | |
| | | | 6 490.06 | 1.81 | 0.397 | 0.16400 | |
| | | | 6 488.83 | 2.81 | 0.176 | 0.11400 | |
| 26 | Fe | K α | 6 404.148 | 1.613 | 0.697 | 0.27800 | [empirical] 7 Lorentzian model from Hölzer et al. (1997). |
| | | | 6 403.295 | 1.965 | 0.376 | 0.18200 | |
| | | | 6 400.653 | 4.833 | 0.088 | 0.10600 | |
| | | | 6 402.077 | 2.803 | 0.136 | 0.09400 | |
| | | | 6 391.190 | 2.487 | 0.339 | 0.20700 | |
| | | | 6 389.106 | 2.339 | 0.060 | 0.06600 | |
| | | | 6 390.275 | 4.433 | 0.102 | 0.06500 | |
| | | K β | 7 046.90 | 14.17 | 0.107 | 0.30100 | [empirical] 4 Lorentzian model from Hölzer et al. (1997). |
| | | | 7 057.21 | 3.12 | 0.448 | 0.27900 | |
| | | | 7 058.36 | 1.97 | 0.615 | 0.24100 | |
| | | | 7 054.75 | 6.38 | 0.141 | 0.17900 | |
| 27 | Co | K α | 6 930.425 | 1.795 | 0.809 | 0.37800 | [empirical] 7 Lorentzian model from Hölzer et al. (1997). |
| | | | 6 929.388 | 2.695 | 0.205 | 0.14400 | |
| | | | 6 927.676 | 4.555 | 0.107 | 0.12700 | |
| | | | 6 930.941 | 0.808 | 0.041 | 0.08800 | |
| | | | 6 915.713 | 2.406 | 0.314 | 0.19700 | |
| | | | 6 914.659 | 2.773 | 0.131 | 0.09500 | |
| | | | 6 913.078 | 4.463 | 0.043 | 0.05000 | |
| | | K β | 7 649.60 | 3.05 | 0.798 | 0.44900 | [empirical] 6 Lorentzian model from Hölzer et al. (1997). |
| | | | 7 647.83 | 3.58 | 0.286 | 0.18900 | |
| | | | 7 639.87 | 9.78 | 0.085 | 0.15300 | |
| | | | 7 645.49 | 4.89 | 0.114 | 0.10300 | |
| | | | 7 636.21 | 13.59 | 0.033 | 0.08200 | |
| | | | 7 654.13 | 3.79 | 0.035 | 0.02500 | |
| 28 | Ni | K α | 7 478.281 | 2.013 | 0.909 | 0.48700 | [empirical] 5 Lorentzian model from Hölzer et al. (1997). |
| | | | 7 476.529 | 4.711 | 0.136 | 0.17100 | |
| | | | 7 461.131 | 2.674 | 0.351 | 0.25000 | |
| | | | 7 459.874 | 3.039 | 0.079 | 0.06400 | |
| | | | 7 458.029 | 4.476 | 0.024 | 0.02800 | |
| | | K β | 8 265.01 | 3.76 | 0.722 | 0.45000 | [empirical] 4 Lorentzian model from Hölzer et al. (1997). |
| | | | 8 263.01 | 4.34 | 0.358 | 0.25800 | |
| | | | 8 256.67 | 13.70 | 0.089 | 0.20300 | |
| | | | 8 268.70 | 5.18 | 0.104 | 0.08900 | |
| 29 | Cu | K α | 8 047.837 | 2.285 | 0.957 | 0.57900 | [empirical] 4 Lorentzian model from Hölzer et al. (1997). |
| | | | 8 045.367 | 3.358 | 0.090 | 0.08000 | |
| | | | 8 027.993 | 2.666 | 0.334 | 0.23600 | |
| | | | 8 026.504 | 3.571 | 0.111 | 0.10500 | |
| | | K β | 8 905.532 | 3.52 | 0.757 | 0.48500 | [empirical] 5 Lorentzian model from Hölzer et al. (1997). |
| | | | 8 903.109 | 3.52 | 0.388 | 0.24800 | |
| | | | 8 908.462 | 3.55 | 0.171 | 0.11000 | |
| | | | 8 897.387 | 8.08 | 0.068 | 0.10000 | |
| | | | 8 911.393 | 5.31 | 0.055 | 0.05500 | |
| 30 | Zn | K α | 8 638.96 | 2.42 | 1.0000 | 0.59780 | [empirical] Ito et al. (2015). Widths of satellite lines are uncorrected. The first and third lines use the CR widths values from Table 1 and the second and fourth use the W widths. |
| | | | 8 636.28 | 4.73 | 0.0602 | 0.07030 | |
| | | | 8 615.90 | 2.45 | 0.4470 | 0.27074 | |
| | | | 8 613.98 | 3.19 | 0.0776 | 0.06115 | |
| 31 | Ga | K α | 9 251.74 | 2.59 | 1.000000 | 0.66032 | [theory] Positions from Bearden (1967) in Zschornack (2007). 2 Lorentzian model (widths) from Krause & Oliver (1979). Relative intensities from Scofield (1974b). |
| | | | 9 224.82 | 2.66 | 0.500889 | 0.33968 | |
| 32 | Ge | K α | 9 886.47 | 2.840 | 1.000 | 0.63650 | [empirical] Ito et al. (2016). Using FWHM column (uncorrected widths) for the satellite lines (2,4) and the CF column (corrected widths) for lines 1 and 3. |
| | | | 9 882.68 | 3.68 | 0.026 | 0.02158 | |
| | | | 9 855.32 | 2.824 | 0.508 | 0.32169 | |
| | | | 9 852.73 | 3.26 | 0.028 | 0.02024 | |
| 33 | As | K α | 10 543.72 | 3.08 | 1.000000 | 0.65994 | [theory] Positions from Bearden (1967). 2 Lorentzian model (widths) from Krause & Oliver (1979). Relative intensities from Scofield (1974a). |
| | | | 10 507.99 | 3.17 | 0.500670 | 0.34006 | |
| 34 | Se | K α | 11 222.4 | 3.33 | 1.000000 | 0.65972 | [theory] Positions from Bearden (1967). 2 Lorentzian model (widths) from Krause & Oliver (1979). Relative intensities from Scofield (1974a). |
| | | | 11 181.4 | 3.46 | 0.496420 | 0.34028 | |
| 35 | Br | K α | 11 924.2 | 3.60 | 1.000000 | 0.65872 | [theory] Positions from Bearden (1967). 2 Lorentzian model (widths) from Krause & Oliver (1979). Relative intensities from Scofield (1974a). |
| | | | 11 877.6 | 3.73 | 0.500043 | 0.34128 | |

References:

- Bearden 1967, Rev. Modern Physics 39, 78 — doi:10.1103/RevModPhys.39.78 — https://ui.adsabs.harvard.edu/abs/1967RvMP...39...78B
- Chantler et al. 2006, PRA 73, 012508 — doi:10.1103/PhysRevA.73.012508 — https://ui.adsabs.harvard.edu/abs/2006PhRvA..73a2508C
- Fischer & Baun 1965, J. Applied Physics 36, 534 — doi:10.1063/1.1714025 — https://ui.adsabs.harvard.edu/abs/1965JAP....36..534F
- Hölzer et al. 1997, PRA 56, 4554 — doi:10.1103/PhysRevA.56.4554 — http://adsabs.harvard.edu/abs/1997PhRvA..56.4554H
- Ito et al. 2015, JQSRT 151, 295 — doi:10.1016/j.jqsrt.2014.10.013 — https://ui.adsabs.harvard.edu/abs/2015JQSRT.151..295I
- Ito et al. 2016, PRA 94, 042506 — doi:10.1103/PhysRevA.94.042506 — https://ui.adsabs.harvard.edu/abs/2016PhRvA..94d2506I
- Kavcic et al. 2005, X-ray Spectrometry 34, 310 — doi:10.1002/xrs.822 — https://ui.adsabs.harvard.edu/abs/2005XRS....34..310K
- Kawai et al. 1994, Spectrochimica Acta B 49, 725 — doi:10.1016/0584-8547(94)80064-2 — https://ui.adsabs.harvard.edu/abs/1994AcSpB..49..725K
- Krause & Oliver 1979, JPCRD 8, 329 — doi:10.1063/1.555595 — http://adsabs.harvard.edu/abs/1979JPCRD...8..329K
- Nordfors 1955, Phys. Soc. A 68, 654 — doi:10.1088/0370-1298/68/7/416 — https://ui.adsabs.harvard.edu/abs/1955PPSA...68..654N
- Schweppe et al. 1994, JESRP 67, 463 doi:10.1016/0368-2048(93)02059-U — https://ui.adsabs.harvard.edu/abs/1994JESRP..67..463S
- Scofield 1974a, PRA 9, 1041 — doi:10.1103/PhysRevA.9.1041 — https://ui.adsabs.harvard.edu/abs/1974PhRvA...9.1041S
- Scofield 1974b, ADNDT 14, 121 — doi:10.1016/S0092-640X(74)80019-7 — https://ui.adsabs.harvard.edu/abs/1974ADNDT..14..121S
- Wollman et al. 2000, NIMPRA 444, 145 — 10.1016/S0168-9002(99)01351-0 — https://ui.adsabs.harvard.edu/abs/2000NIMPA.444..145W
- Zschornack 2007, Handbook of X-ray Data,Springer — doi:10.1007/978-3-540-28619-6 — https://link.springer.com/book/10.1007/978-3-540-28619-6

Compiled by the NASA/GSFC calorimeter group (contact: F.S. Porter, frederick.s.porter@nasa.gov ; M.C. Witthoeft, michael.c.witthoeft@nasa.gov ).
Cite the quoted references when using these values!

Table 4: E models (measurement, theory, or combination) for lineshape and position of K-shell lines in neutrals (here: solid targets). Listed are energy (eV), Lorentzian width Γ (FWHM in eV), and relative amplitude (area $\cdot 2/\pi/\Gamma$) of the model components. Also listed are the normalized intensities (area) of the components with $\sum_i I_i = 1$. Note that the exact line shapes depend on source conditions (solid, gas; molecules causing chemical shifts), excitation process, and excitation energy.

| $Z$ | | Trans. Type | Energy [eV] | FWHM Γ [eV] | Amplitude area $\cdot 2/\pi/\Gamma$ | Intensity (normalized) | References |
|---|---|---|---|---|---|---|---|
| 37 | Rb | K α | 13 395.3<br>13 335.8 | 4.42<br>4.26 | 1.000000<br>0.539012 | 0.65811<br>0.34189 | [theory] Positions from Bearden (1967). 2 Lorentzian model (widths) from Krause & Oliver (1979). Note: the K$\alpha_1$ width of 4.92 eV quoted in the paper is a clear outlier to the $Z$-trend; adjusting the value to 4.42 eV (summed widths for K and $L_2$ levels) fits the trend well. Relative intensities from Scofield (1974a). |
| 39 | Y | K α | 14 958.54<br>14 882.94 | 5.02<br>5.18 | 1.000000<br>0.505718 | 0.65710<br>0.34290 | [theory] Positions from Bearden (1967) in Zschornack (2007). 2 Lorentzian model (widths) from Krause & Oliver (1979). Relative intensities from Scofield (1974b). |
| 40 | Zr | K α | 15 775.1<br>15 690.9 | 5.40<br>5.62 | 1.000000<br>0.502046 | 0.65681<br>0.34319 | [theory] Positions from Bearden (1967). 2 Lorentzian model (widths) from Krause & Oliver (1979). Relative intensities from Scofield (1974a). |
| 41 | Nb | K α | 16 615.1<br>16 521.0 | 5.80<br>6.01 | 1.000000<br>0.505369 | 0.65631<br>0.34369 | [theory] Positions from Bearden (1967). 2 Lorentzian model (widths) from Krause & Oliver (1979). Relative intensities from Scofield (1974b). |
| 42 | Mo | K α | 17 479.34<br>17 374.3 | 6.31<br>6.49 | 1.000000<br>0.510147 | 0.65587<br>0.34413 | [theory] Positions from Bearden (1967). 2 Lorentzian model (widths) from Krause & Oliver (1979). Relative intensities from Scofield (1974a). |
| 47 | Ag | K α | 22 162.92<br>21 990.3 | 9.16<br>9.32 | 1.000000<br>0.521393 | 0.65338<br>0.34662 | [theory] Positions from Bearden (1967). 2 Lorentzian model (widths) from Krause & Oliver (1979). Relative intensities from Scofield (1974a). |
| 50 | Sn | K α | 25 271.3<br>25 044.0 | 11.2<br>11.3 | 1.000000<br>0.529572 | 0.65176<br>0.34824 | [theory] Positions from Bearden (1967). 2 Lorentzian model (widths) from Krause & Oliver (1979). Relative intensities from Scofield (1974a). |
| | | K β | 28 486.0<br>28 444.0 | 11.8<br>11.0 | 1.000000<br>0.552240 | 0.66015<br>0.33985 | [theory] Positions from Bearden (1967). Relative intensities from Scofield (1974a). |

References:

- Bearden 1967, Rev. Modern Physics 39, 78 — doi:10.1103/RevModPhys.39.78 — https://ui.adsabs.harvard.edu/abs/1967RvMP...39...78B
- Chantler et al. 2006, PRA 73, 012508 — doi:10.1103/PhysRevA.73.012508 — https://ui.adsabs.harvard.edu/abs/2006PhRvA..73a2508C
- Fischer & Baun 1965, J. Applied Physics 36, 534 — doi:10.1063/1.1714025 — https://ui.adsabs.harvard.edu/abs/1965JAP....36..534F
- Hölzer et al. 1997, PRA 56, 4554 — doi:10.1103/PhysRevA.56.4554 — http://adsabs.harvard.edu/abs/1997PhRvA..56.4554H
- Ito et al. 2015, JQSRT 151, 295 — doi:10.1016/j.jqsrt.2014.10.013 — https://ui.adsabs.harvard.edu/abs/2015JQSRT.151..295I
- Ito et al. 2016, PRA 94, 042506 — doi:10.1103/PhysRevA.94.042506 — https://ui.adsabs.harvard.edu/abs/2016PhRvA..94d2506I
- Kavcic et al. 2005, X-ray Spectrometry 34, 310 — doi:10.1002/xrs.822 — https://ui.adsabs.harvard.edu/abs/2005XRS....34..310K
- Kawai et al. 1994, Spectrochimica Acta B 49, 725 — doi:10.1016/0584-8547(94)80064-2 — https://ui.adsabs.harvard.edu/abs/1994AcSpB..49..725K
- Krause & Oliver 1979, JPCRD 8, 329 — doi:10.1063/1.555595 — http://adsabs.harvard.edu/abs/1979JPCRD...8..329K
- Nordfors 1955, Phys. Soc. A 68, 654 — doi:10.1088/0370-1298/68/7/416 — https://ui.adsabs.harvard.edu/abs/1955PPSA...68..654N
- Schweppe et al. 1994, JESRP 67, 463 doi:10.1016/0368-2048(93)02059-U — https://ui.adsabs.harvard.edu/abs/1994JESRP..67..463S
- Scofield 1974a, PRA 9, 1041 — doi:10.1103/PhysRevA.9.1041 — https://ui.adsabs.harvard.edu/abs/1974PhRvA...9.1041S
- Scofield 1974b, ADNDT 14, 121 — doi:10.1016/S0092-640X(74)80019-7 — https://ui.adsabs.harvard.edu/abs/1974ADNDT..14..121S
- Wollman et al. 2000, NIMPRA 444, 145 — 10.1016/S0168-9002(99)01351-0 — https://ui.adsabs.harvard.edu/abs/2000NIMPA.444..145W
- Zschornack 2007, Handbook of X-ray Data,Springer — doi:10.1007/978-3-540-28619-6 — https://link.springer.com/book/10.1007/978-3-540-28619-6

Compiled by the NASA/GSFC calorimeter group (contact: F.S. Porter, frederick.s.porter@nasa.gov ; M.C. Witthoeft, michael.c.witthoeft@nasa.gov ).

Cite the quoted references when using these values!